\documentclass[conference]{IEEEtran}
\usepackage{algorithmic}
\usepackage{amsmath,amssymb,amsfonts}
\usepackage{tikz}
\usepackage{caption}
\usepackage{cite}
\usepackage{subcaption}
\usepackage{graphicx}
\usetikzlibrary{arrows.meta, positioning}
\usepackage{pgfplots}
\pgfplotsset{compat=newest}
\usetikzlibrary{positioning}
\usepackage{float}
\usepackage{comment}
\pgfplotsset{compat=1.18}
\usepgfplotslibrary{groupplots}
\usepackage{pgfplotstable}
\usepackage{xcolor}
\usetikzlibrary{patterns}
\usepackage{rotating}
\usepackage{multirow}
\usepackage{amsmath}
\usetikzlibrary{shapes.geometric, arrows.meta, positioning, fit, backgrounds, calc, shadows.blur}
\usepackage{helvet}
\usepackage{times}
\usepackage{url}

\begin{document}
\title{Surrogates, Spikes, and Sparsity: Performance Analysis and Characterization of SNN Hyperparameters on Hardware}

\author{Ilkin Aliyev, Jesus Lopez, and Tosiron Adegbija\\
Department of Electrical and Computer Engineering\\
The University of Arizona, Tucson, AZ, USA\\
Email: \{ilkina, jlopezramos, tosiron\}@arizona.edu
}

\maketitle

\begin{abstract}

Spiking Neural Networks (SNNs) offer inherent advantages for low-power inference through sparse, event-driven computation. However, the theoretical energy benefits of SNNs are often decoupled from real-world hardware performance due to the opaque relationship between training-time choices and inference-time sparsity. While prior work has focused on weight pruning and model compression, the role of training hyperparameters---specifically surrogate gradient functions and neuron model configurations---in shaping hardware-level activation sparsity remains underexplored. 

This paper presents a comprehensive workload characterization study quantifying the sensitivity of hardware latency to SNN hyperparameters. We decouple the impact of surrogate gradient functions (e.g., Fast Sigmoid, Spike Rate Escape) and neuron models (LIF, Lapicque) on classification accuracy and inference efficiency across three event-based vision datasets: DVS128-Gesture, N-MNIST, and DVS-CIFAR10. Our analysis reveals that standard accuracy metrics are poor predictors of hardware efficiency. For instance, while Fast Sigmoid achieves the highest accuracy on DVS-CIFAR10, the Spike Rate Escape reduces inference latency by up to 12.2\% on DVS128-Gesture with minimal accuracy trade-offs. Furthermore, we demonstrate that neuron model selection is as critical as parameter tuning; transitioning from LIF to Lapicque neurons yields up to a 28\% latency reduction. We validate our analysis on a custom cycle-accurate FPGA-based SNN instrumentation platform, and our characterization demonstrates that sparsity-aware hyperparameter selection can improve accuracy by 9.1\% and latency by over 2$\times$ compared to baselines. These findings establish a methodology for predicting hardware behavior from training parameters, motivating the inclusion of sparsity-sensitivity in future SNN performance analysis. The RTL code and other reproducibility artifacts are available at \url{https://zenodo.org/records/18893738}.
\end{abstract}

\section{Introduction}

Spiking Neural Networks (SNNs) have garnered significant attention as a biologically plausible and energy-efficient alternative to traditional Artificial Neural Networks (ANNs). Unlike ANNs, which process information using dense and power-intensive matrix multiplications, SNNs leverage sparse event-driven computations where neurons activate only when a membrane potential threshold is reached. This paradigm promises efficient temporal information encoding and reduced energy consumption, making SNNs highly attractive for neuromorphic hardware and edge AI applications \cite{yin2022sata, davies2021loihi, zhang2024spiking, liu2024energy}.

However, a critical gap remains between the theoretical benefits of SNNs and their performance on hardware. While the efficiency of SNNs is predicated on activation sparsity \cite{bouvier2019spiking,nunes2022spiking}, this sparsity is not a static property of the network; it is dynamically shaped by training-time hyperparameters that are often selected solely for classification accuracy. Consequently, widespread adoption is hindered not only by the challenge of training non-differentiable spikes \cite{neftci2019surrogate} but also by the opacity of how algorithmic choices---specifically, surrogate gradient functions and neuron models---impact downstream inference latency and efficiency. This gap is particularly critical for event-based workloads, where temporal sparsity is intrinsic to the input modality rather than an artifact of encoding. While weight pruning has been extensively characterized \cite{yin2024workload, wang2023spiking}, the hardware sensitivity of SNNs to activation-governing hyperparameters remains an underexplored domain.

The primary source of this disconnect lies in the \textit{surrogate gradient function} used to approximate non-differentiable spikes during backpropagation. These functions are mathematical abstractions necessary for learning, yet they subtly dictate the spiking intensity (i.e., firing rate) of the trained model. Different approximations, such as \texttt{Fast Sigmoid} or \texttt{Arctangent}, impose varying gradients that result in distinct sparsity profiles. Recent work \cite{aliyev2024fine} has begun to examine these functions, but a systematic characterization of their hardware implications, and how they trade off accuracy for latency across different workloads, is lacking \cite{shrestha2018slayer, yin2022sata}.

Furthermore, the choice of the neuron model itself (e.g., Leaky Integrate-and-Fire vs. Lapicque) and its internal parameters (decay $\beta$, threshold $\theta$) introduces another layer of complexity. Current methodologies often rely on fixed or heuristic-based configurations, neglecting the workload-dependent nature of optimal activation dynamics. This is particularly crucial for event-based vision workloads (e.g., DVS gestures), where temporal sparsity is the primary driver of efficiency. Without a rigorous characterization of these factors, SNNs risk being deployed with ``accuracy-optimized" configurations that are computationally sub-optimal on real hardware.

\begin{figure*}[!ht]
    \centering
    \resizebox{0.8\linewidth}{!}{\begin{tikzpicture}[
    font=\footnotesize,
    node distance=1.4cm,
    >={Stealth[length=2mm, width=2.5mm]},
    opbox/.style={
        rectangle,
        rounded corners=3pt,
        draw=black,
        line width=0.5pt,
        fill=#1,
        minimum width=1.8cm,
        minimum height=0.7cm,
        align=center,
        font=\footnotesize
    },
    databox/.style={
        rectangle,
        rounded corners=2pt,
        draw=gray!60,
        line width=0.4pt,
        fill=gray!10,
        minimum width=1.4cm,
        minimum height=0.45cm,
        align=center,
        font=\scriptsize
    },
    section/.style={
        rectangle,
        rounded corners=5pt,
        draw=#1,
        line width=1pt,
        inner sep=8pt
    },
    arrow/.style={
        ->,
        line width=0.5pt
    },
    cyclebar/.style={
        rectangle,
        draw=none,
        fill=#1,
        minimum height=0.35cm
    }
]

\definecolor{lifblue}{RGB}{70,130,180}
\definecolor{lapgreen}{RGB}{60,140,100}
\definecolor{addcolor}{RGB}{200,220,240}
\definecolor{multcolor}{RGB}{255,220,180}
\definecolor{shiftcolor}{RGB}{220,240,220}
\definecolor{cmpcolor}{RGB}{240,220,240}


\node[databox] (lif_mem) {$U[t-1]$};
\node[databox, right=1.2cm of lif_mem] (lif_in) {$I_{syn}[t]$};

\node[opbox=shiftcolor, below=0.5cm of lif_mem] (lif_shift) {\textbf{Bit-Shift}};
\node[font=\scriptsize, gray, left=0.1cm of lif_shift, anchor=east](beta_text) {$\beta \cdot U$};

\node[opbox=addcolor, below=0.5cm of lif_shift] (lif_add) {\textbf{Add}};

\node[opbox=cmpcolor, below=0.5cm of lif_add] (lif_cmp) {\textbf{Compare}};
\node[font=\scriptsize, gray, left=0.1cm of lif_cmp, anchor=east] {$U > \theta$};

\node[databox, below=0.5cm of lif_cmp] (lif_out) {Spike / No Spike};

\draw[arrow] (lif_mem) -- (lif_shift);
\draw[arrow] (lif_shift) -- (lif_add);
\draw[arrow] (lif_in) -- ++(0,-0.25) to[out=-90, in=0](lif_add.east);
\draw[arrow] (lif_add) -- (lif_cmp);
\draw[arrow] (lif_cmp) -- (lif_out);

\begin{scope}[on background layer]
    \node[section=lifblue, fill=lifblue!8, fit=(lif_mem)(lif_in)(beta_text)(lif_shift)(lif_add)(lif_cmp)(lif_out), 
          inner xsep=15pt, inner ysep=10pt] (lifbox) {};
\end{scope}

\node[font=\footnotesize\bfseries, lifblue, anchor=south] at (lifbox.north) {LIF Neuron};

\node[font=\scriptsize, lifblue!80!black, anchor=north, align=center] at ([yshift=-0.1cm]lifbox.south) 
    {\textit{Low Arithmetic Intensity}\\[-2pt]\scriptsize (3 ops/update)};


\node[databox, right=6cm of lif_mem] (lap_mem) {$U[t-1]$};
\node[databox, right=1.2cm of lap_mem] (lap_in) {$I_{syn}[t]$};

\node[opbox=multcolor, below=0.5cm of lap_mem] (lap_mult1) {\textbf{Multiply}};
\node[font=\scriptsize, gray, left=0.1cm of lap_mult1, anchor=east] (u_rc){$U \cdot (1{-}\frac{T}{RC})$};

\node[opbox=multcolor, below=0.5cm of lap_in] (lap_mult2) {\textbf{Multiply}};
\node[font=\scriptsize, gray, right=0.1cm of lap_mult2, anchor=west] (t_rc) {$I \cdot \frac{T}{RC}$};

\node[opbox=addcolor, below=0.8cm of lap_mult1, xshift=1.35cm] (lap_add) {\textbf{Add}};

\node[opbox=cmpcolor, below=0.5cm of lap_add] (lap_cmp) {\textbf{Compare}};
\node[font=\scriptsize, gray, left=0.1cm of lap_cmp, anchor=east] {$U > \theta$};

\node[databox, below=0.5cm of lap_cmp] (lap_out) {Spike / No Spike};

\draw[arrow] (lap_mem) -- (lap_mult1);
\draw[arrow] (lap_in) -- (lap_mult2);
\draw[arrow] (lap_mult1) to[out=-90, in=180] (lap_add.west);
\draw[arrow] (lap_mult2) to[out=-90, in=0] (lap_add.east);
\draw[arrow] (lap_add) -- (lap_cmp);
\draw[arrow] (lap_cmp) -- (lap_out);

\begin{scope}[on background layer]
    \node[section=lapgreen, fill=lapgreen!8, fit=(lap_mem)(lap_in)(u_rc)(lap_mult1)(lap_mult2)(lap_add)(lap_cmp)(lap_out)(t_rc), 
          inner xsep=15pt, inner ysep=10pt] (lapbox) {};
\end{scope}

\node[font=\footnotesize\bfseries, lapgreen, anchor=south] at (lapbox.north) {Lapicque Neuron};

\node[font=\scriptsize, lapgreen!80!black, anchor=north, align=center] at ([yshift=-0.1cm]lapbox.south) 
    {\textit{High Arithmetic Intensity}\\[-2pt]\scriptsize (4 ops/update, 2$\times$ MUL)};


\coordinate (barstart) at ([yshift=-1.2cm, xshift=-0.5cm]lifbox.south west);

\node[anchor=east, font=\scriptsize] at ([yshift=-0.55cm, xshift=1.2cm]barstart) {LIF:};

\fill[lifblue!50] ([yshift=-0.7cm, xshift=1.3cm]barstart) rectangle ++(1.0cm, 0.3cm);
\fill[lifblue!50] ([yshift=-0.7cm, xshift=2.5cm]barstart) rectangle ++(1.0cm, 0.3cm);
\fill[lifblue!50] ([yshift=-0.7cm, xshift=3.7cm]barstart) rectangle ++(1.0cm, 0.3cm);
\fill[lifblue!50] ([yshift=-0.7cm, xshift=4.9cm]barstart) rectangle ++(1.0cm, 0.3cm);
\fill[lifblue!50] ([yshift=-0.7cm, xshift=6.1cm]barstart) rectangle ++(1.0cm, 0.3cm);

\node[anchor=east, font=\scriptsize] at ([yshift=-1.35cm, xshift=1.2cm]barstart) {Lapicque:};

\fill[lapgreen!50] ([yshift=-1.5cm, xshift=1.3cm]barstart) rectangle ++(1.6cm, 0.3cm);
\fill[lapgreen!50] ([yshift=-1.5cm, xshift=4.5cm]barstart) rectangle ++(1.6cm, 0.3cm);

\draw[decorate, decoration={brace, amplitude=3pt, raise=2pt}, line width=0.4pt, lifblue] 
    ([yshift=-0.4cm, xshift=1.3cm]barstart) -- node[above=4pt, font=\scriptsize, lifblue!70!black] {5 updates (frequent)} ([yshift=-0.4cm, xshift=7.1cm]barstart);
    
\draw[decorate, decoration={brace, amplitude=3pt, mirror, raise=2pt}, line width=0.4pt, lapgreen] 
    ([yshift=-1.5cm, xshift=1.3cm]barstart) -- node[below=4pt, font=\scriptsize, lapgreen!70!black] {2 updates (sparser activations)} ([yshift=-1.5cm, xshift=6.1cm]barstart);

\draw[->, line width=0.4pt] ([yshift=-2.0cm, xshift=1.3cm]barstart) -- ([yshift=-2.0cm, xshift=8.5cm]barstart);
\node[font=\scriptsize, anchor=west] at ([yshift=-2.0cm, xshift=8.6cm]barstart) {Time};

\node[draw=gray!60, rounded corners=3pt, fill=yellow!10, inner sep=5pt, font=\scriptsize, align=left, anchor=north west] 
    at ([yshift=-2.4cm, xshift=0cm]barstart) {
    Lapicque has higher \textit{per-update} cost (2$\times$ multiply), but triggers \textbf{fewer updates} due to sparser activations $\rightarrow$ \textbf{lower total latency}.
};

\end{tikzpicture}}  
    \caption{Hardware dataflow comparison between the LIF and Lapicque neuron models. LIF uses bit-shifts for decay (3 ops/update); Lapicque requires explicit multiplications for RC constants (4 ops + 2$\times$ MUL). Despite higher per-update cost, Lapicque's temporal dynamics suppress total spike events, yielding lower system-level latency.}
    \label{fig:neuron_dataflow}
\end{figure*}
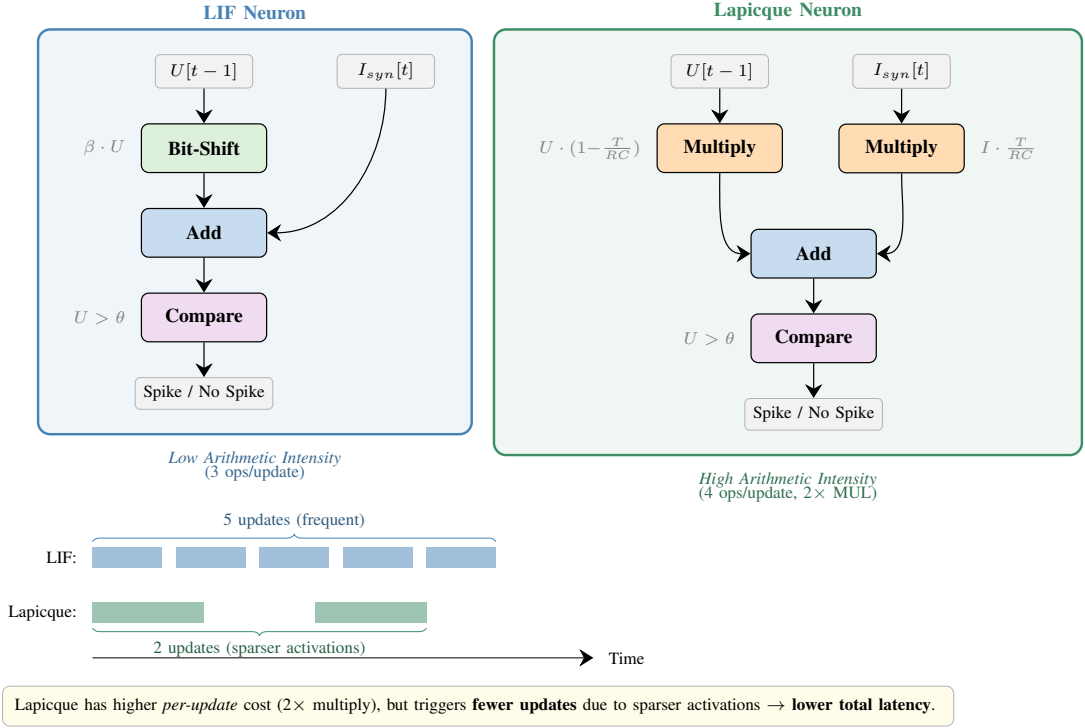

This paper presents a systematic \textbf{workload characterization study} investigating the interplay between training-time hyperparameters and inference-time hardware performance. We move beyond simple accuracy metrics to quantify the latency and sparsity sensitivity of SNNs on a cycle-accurate hardware platform. The key contributions are:

\begin{itemize}
    \item \textbf{Characterization of surrogate gradient impact:} We isolate and evaluate four distinct surrogate gradient functions---\texttt{Fast Sigmoid (FS)}, \texttt{Arctangent (ATAN)}, \texttt{Spike Rate Escape (SRE)}, and \texttt{Stochastic Spike Operator (SSO)}---quantifying their non-trivial trade-offs between classification accuracy and hardware activation sparsity.
    \item \textbf{Analysis of neuron model sensitivity:} We characterize the performance implications of neuron model selection, contrasting the standard \textit{Leaky Integrate-and-Fire (LIF)} model with the biologically grounded \textit{Lapicque (LAP)} model. We demonstrate that switching neuron models can yield up to a \textbf{28\% reduction in latency} through improved sparsity dynamics, a factor often overlooked in software-centric optimizations.
    \item \textbf{Workload-specific sparsity profiling:} We focus on three dynamic event-based datasets: \textit{DVS128-Gesture}, \textit{N-MNIST}, and \textit{DVS-CIFAR10}. Our analysis reveals that, unlike static image workloads, these event streams exhibit unique sparsity characteristics that require distinct hyperparameter configurations to maximize efficiency.
    \item \textbf{Hardware-in-the-loop validation:} To ensure our characterization reflects real-world constraints, we validate our findings on a custom sparsity-aware FPGA instrumentation platform. We show that our methodology identifies configurations that achieve \textbf{9.1\% higher accuracy} and over \textbf{2$\times$ lower latency} compared to prior hardware baselines \cite{sommer2022efficient,yin2022sata}, proving that training-time hyperparameters are a first-order determinant of hardware performance.
\end{itemize}

\section{Background} \label{sec:background}

\subsection{Spiking Neuron Models as Hardware Workloads}

Spiking Neural Networks (SNNs) fundamentally differ from ANNs by introducing stateful temporal dynamics. From a hardware perspective, this shifts the computational bottleneck from dense matrix multiplication to state updates and sparse event processing. The complexity of the workload is governed by the specific neuron model employed. We analyze two distinct models:

\noindent\textbf{Leaky Integrate-and-Fire (LIF):} The LIF model is the standard for hardware-efficient SNNs due to its low arithmetic intensity \cite{gerstner2014neuronal}. The membrane potential $u_j[t]$ of neuron $j$ accumulates input from presynaptic spikes $s_i[t] \in \{0,1\}$, scaled by synaptic weights $w_{ij}$, and decays by a factor $\beta \in (0,1)$:

\begin{equation} \label{eq:lif}
    u_{j}[t+1] = \beta u_{j}[t] + \sum_{i} w_{ij} s_{i}[t] - s_{j}[t] \theta
\end{equation}

\noindent where  $\theta$ is the firing threshold and the term $s_{j}[t]\theta$ resets the membrane upon spike emission. This formulation requires only simple accumulation and shifting (if $\beta$ is a power of two), making it attractive for digital logic.

\noindent\textbf{Lapicque (LAP):} The Lapicque model \cite{brunel07_lapicque} introduces higher biological fidelity by modeling the membrane as an $RC$ circuit. The potential decays exponentially with a time constant $\tau = RC$. While the LIF model often abstracts this decay using a fixed parameter $\beta$, the LAP model retains explicit control over $R$ and $C$, offering a more biologically plausible and tunable formulation. Its discrete formulation is:

\begin{equation} \label{eq:lap}
    u_{j}[t+1] = \left(1 - \frac{T}{RC}\right) u_{j}[t] + \sum_{i} w_{ij} s_{i}[t] \left(\frac{T}{C}\right) - s_{j}[t] \theta
\end{equation}

In our framework, we map the decay parameters to an equivalent capacitance:

\begin{align}
C &= -\frac{1}{\ln(\beta)}
\end{align}

Figure~\ref{fig:neuron_dataflow} contrasts the hardware dataflow of these two models, highlighting the increased arithmetic intensity (specifically the requirement for multipliers) introduced by the biological fidelity of the Lapicque formulation. While LAP implies higher arithmetic complexity (potentially requiring multipliers or lookup tables for the RC constants) compared to a simplified LIF, it offers distinct temporal dynamics that may lead to increased sparsity. A key question of this study is whether the increased per-operation cost of the LAP model is offset by the reduced total operations (increased sparsity) during inference.

\begin{figure*}[!ht]
    \centering
    \resizebox{0.7\linewidth}{!}{\begin{tikzpicture}

\definecolor{fscolor}{RGB}{31,119,180}    
\definecolor{atancolor}{RGB}{255,127,14}  
\definecolor{srecolor}{RGB}{44,160,44}    
\definecolor{ssocolor}{RGB}{148,103,189}  

\begin{axis}[
    name=plot1,
    at={(0,0)},
    width=5.2cm, height=4.5cm,
    xlabel={$U - U_{\text{thr}}$}, ylabel={Gradient Magnitude},
    xlabel style={font=\footnotesize}, ylabel style={font=\footnotesize},
    tick label style={font=\scriptsize},
    xmin=-3, xmax=3, ymin=0, ymax=1.1,
    axis lines=left, grid=major, grid style={gray!30},
    legend style={
        at={(1.0,0.98)}, anchor=north east, font=\tiny, 
        draw=gray!50, fill=white, fill opacity=0.95,
        rounded corners=1pt, inner sep=2pt, row sep=-2pt
    },
    clip=true
]
    \addplot[fscolor!50, line width=1pt, samples=100, domain=-3:3] {1 / ((1 + 1*abs(x))^2)};
    \addplot[fscolor!75, line width=1.1pt, samples=100, domain=-3:3] {1 / ((1 + 2.5*abs(x))^2)};
    \addplot[fscolor, line width=1.3pt, samples=100, domain=-3:3] {1 / ((1 + 5*abs(x))^2)};
    \legend{$k=1$, $k=2.5$, $k=5$}
\end{axis}
\node[below=25pt of plot1.south, font=\footnotesize] {(a) Fast Sigmoid (FS)};

\begin{axis}[
    name=plot2,
    at={(5.5cm,0)},
    width=5.2cm, height=4.5cm,
    xlabel={$U - U_{\text{thr}}$}, ylabel={},
    xlabel style={font=\footnotesize},
    tick label style={font=\scriptsize},
    xmin=-3, xmax=3, ymin=0, ymax=0.4,
    axis lines=left, grid=major, grid style={gray!30},
    legend style={
        at={(1.0,0.98)}, anchor=north east, font=\tiny, 
        draw=gray!50, fill=white, fill opacity=0.95,
        rounded corners=1pt, inner sep=2pt, row sep=-2pt
    },
    clip=true
]
    \addplot[atancolor!50, line width=1pt, samples=100, domain=-3:3] {1 / (3.14159 * (1 + (3.14159*x*1/2)^2))};
    \addplot[atancolor!75, line width=1.1pt, samples=100, domain=-3:3] {1 / (3.14159 * (1 + (3.14159*x*2.5/2)^2))};
    \addplot[atancolor, line width=1.3pt, samples=100, domain=-3:3] {1 / (3.14159 * (1 + (3.14159*x*5/2)^2))};
    \legend{$\alpha=1$, $\alpha=2.5$, $\alpha=5$}
\end{axis}
\node[below=25pt of plot2.south, font=\footnotesize] {(b) Arctangent (ATAN)};

\begin{axis}[
    name=plot3,
    at={(0,-5.2cm)},
    width=5.2cm, height=4.5cm,
    xlabel={$U - U_{\text{thr}}$}, ylabel={Gradient Magnitude},
    xlabel style={font=\footnotesize}, ylabel style={font=\footnotesize},
    tick label style={font=\scriptsize},
    xmin=-3, xmax=3, ymin=0, ymax=5.5,
    axis lines=left, grid=major, grid style={gray!30},
    legend style={
        at={(1.0,0.98)}, anchor=north east, font=\tiny, 
        draw=gray!50, fill=white, fill opacity=0.95,
        rounded corners=1pt, inner sep=2pt, row sep=-2pt
    },
    clip=true
]
    \addplot[srecolor!60, line width=1.1pt, samples=200, domain=-3:3] {1 * exp(-1*abs(x))};
    \addplot[srecolor!80, line width=1.2pt, samples=200, domain=-3:3] {2.5 * exp(-2.5*abs(x))};
    \addplot[srecolor, line width=1.4pt, samples=200, domain=-3:3] {5 * exp(-5*abs(x))};
    \legend{$\beta=1$, $\beta=2.5$, $\beta=5$}
\end{axis}
\node[below=25pt of plot3.south, font=\footnotesize] {(c) Spike Rate Escape (SRE)};

\begin{axis}[
    name=plot4,
    at={(5.5cm,-5.2cm)},
    width=5.2cm, height=4.5cm,
    xlabel={$U - U_{\text{thr}}$}, ylabel={},
    xlabel style={font=\footnotesize},
    tick label style={font=\scriptsize},
    xmin=-3, xmax=3, ymin=-0.2, ymax=1.2,
    axis lines=left, grid=major, grid style={gray!30},
    clip=true
]
    \fill[ssocolor!20, opacity=0.5] (axis cs:-3,-0.1) rectangle (axis cs:0,0.1);

    \addplot[ssocolor, line width=1.5pt, domain=0:3, samples=2] {1};

    \addplot[
        only marks,
        mark=*,
        mark size=0.3pt,
        ssocolor!70,
        samples=150,
        domain=-3:0
    ] {(rand - 0.5)*0.2};

    \addplot[ssocolor!50, dashed, thin, domain=-3:0, samples=2] {0.1};
    \addplot[ssocolor!50, dashed, thin, domain=-3:0, samples=2] {-0.1};

    \node[font=\tiny, ssocolor!80!black, align=center]
        at (axis cs:1.05,1.12) {$x \geq 0$ \\ straight-through gradient ($=1$)};

    \node[font=\tiny, ssocolor!80!black, align=center]
        at (axis cs:-1.5,0.25) {$x < 0$ \\ stochastic gradient};

\end{axis}
\node[below=25pt of plot4.south, font=\footnotesize] {(d) Stochastic Spike Operator (SSO)};

\end{tikzpicture}}  
    \caption{Visualization of the derivative profiles ($\frac{\partial S}{\partial U}$) for different surrogate functions plotted against the membrane potential's proximity to the firing threshold ($U - U_{thr}$). Fast Sigmoid (a) maintains meaningful gradient support in the tails as $k$ increases. Arctangent (b) exhibits similar behavior with Cauchy-like tails that preserve gradient flow farther from the threshold. Spike Rate Escape (c) exhibits exponential decay, leading to vanishing gradients (gradient $\approx$ 0) for neurons far from the threshold (assuming $\beta = k$). The Stochastic Spike Operator (d) uses a constant gradient of 1 above threshold and injects uniform noise below threshold, enabling gradient flow through sub-threshold neurons. These mathematical properties explain the sharp accuracy degradation (``cliffs") observed in our experimental characterization (Section~\ref{sec:results}).}
    \label{fig:surrogates}
\end{figure*}
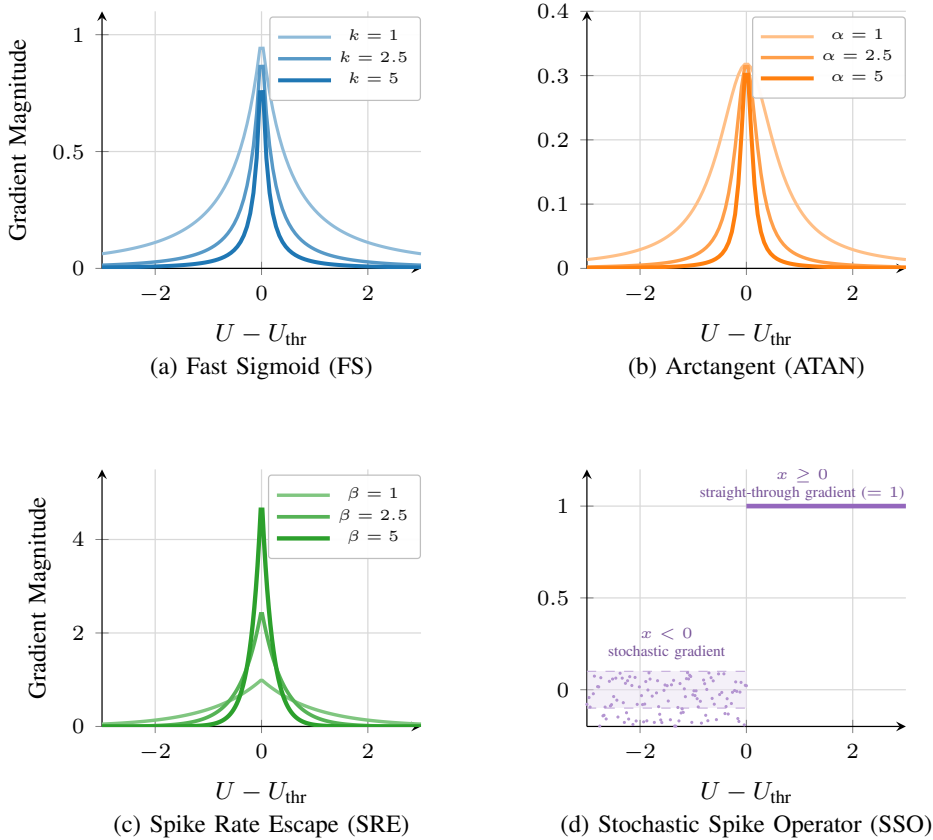

\begin{table*}[t]
    \centering
    \renewcommand{\arraystretch}{1.8}
    \setlength{\tabcolsep}{6pt}
    \caption{Overview of surrogate approximation functions as implemented in snnTorch \cite{eshraghian2023training}. All functions are written in threshold-centered form with $x = U - U_{\text{thr}}$.}
    \begin{tabular}{|l|l|p{7.2cm}|}
        \hline
        \textbf{Surrogate Function} & \textbf{Backward Pass Formula} & \textbf{Key Characteristics} \\ 
        \hline
        \textbf{Fast Sigmoid (FS)} &
        $\displaystyle \frac{\partial S}{\partial U} \approx \frac{1}{(1 + k|x|)^2}$ &
        Default snnTorch surrogate; computationally cheap with a sharp peak at threshold and quadratic heavy-tail decay. Parameter $k$ (``slope'') controls sharpness (larger $k$ $\rightarrow$ narrower peak). \\ 
        \hline
        \textbf{Arctangent (ATAN)} &
        $\displaystyle \frac{\partial S}{\partial U} \approx \frac{1}{\pi}\cdot\frac{1}{1 + \left(\pi x\frac{\alpha}{2}\right)^2}$ &
        Smooth, symmetric gradient with Cauchy-like tails; tends to preserve gradient flow farther from threshold than FS. Parameter $\alpha$ controls sharpness. \\ 
        \hline
        \textbf{Spike Rate Escape (SRE)} &
        $\displaystyle \frac{\partial S}{\partial U} \approx k\,\exp\!\big(-\beta\,|x + (U_{\text{thr}}-1)|\big)$ &
        Escape-rate (Boltzmann-like) exponential surrogate as implemented in snnTorch: centered at $U\!=\!1$. If $U_{\text{thr}}=1$ (common default), this simplifies to $\frac{\partial S}{\partial U}\approx k e^{-\beta|x|}$. Parameters: $k$ sets peak scale; $\beta$ controls decay with distance from threshold. \\ 
        \hline
        \textbf{Stochastic Spike Operator (SSO)} &
        $\displaystyle \frac{\partial S}{\partial U} \approx 
        \begin{cases}
        1 & x \ge 0 \\
        \left(\mathcal{U}(-0.5,\,0.5)+\mu\right)\sigma^{2} & x < 0
        \end{cases}$ &
        Stochastic sub-threshold gradients: above threshold uses straight-through gradient 1; below threshold samples a uniform random value (shifted by mean $\mu$) scaled by variance term $\sigma^2$. Intended to mitigate ``dead'' neurons by injecting stochastic gradient signal when $U<U_{\text{thr}}$. \\ 
        \hline
    \end{tabular}
    \label{table:surrogate_overview}
\end{table*}

\subsection{Surrogate Gradients and Sparsity Determinism}

Backpropagation in SNNs presents a fundamental challenge due to the non-differentiability of the spiking function. Since neurons output discrete binary values (spike or no spike), conventional gradient-based optimization fails to compute meaningful updates. To circumvent this, surrogate gradient techniques approximate the derivative of the non-differentiable spike function using smooth functions \cite{neftci2019surrogate}. While these functions are only active during training, they deterministically shape the weight distribution and firing thresholds, thereby informing the sparsity profile observed during hardware inference.

\subsubsection{Common Surrogate Functions}

Surrogate functions are commonly used during training (implemented, for example, in PyTorch’s \textit{autograd} module) to enable gradient-based training of SNNs. The general form of the surrogate gradient approximation is:

\begin{equation}
    \frac{\partial S}{\partial U} \approx f(U, U_{\text{thr}}, \alpha)
\end{equation}

where $U$ is the membrane potential, $U_{\text{thr}}$ is the firing threshold, and $\alpha$ is a tunable derivative scaling factor that controls the sharpness of the function. We characterize four surrogates (Table~\ref{table:surrogate_overview}) with distinct gradient behaviors:

\begin{itemize}
\item \textbf{Fast Sigmoid (FS)}: A smooth, S-shaped approximation of a step function centered around the middle point. It is simple and fast to compute, making it efficient for large-scale training.

\item \textbf{Arctangent (ATAN)}: Similar in shape to the Fast Sigmoid but slightly more computationally demanding. It uses an arctangent form that also produces a smooth transition but involves more arithmetic operations.

\item \textbf{Spike Rate Escape (SRE)}: An exponentially decaying function that reacts strongly to how close the membrane potential is to the firing threshold. It is highly sensitive and useful for modeling rapid changes near the spiking point.

\item \textbf{Stochastic Spike Operator (SSO)}: A surrogate with stochastic sub-threshold gradients. The forward pass uses the standard Heaviside step function, while the backward pass returns a gradient of 1 above threshold and samples uniform random noise (scaled by variance $\sigma^2$) below threshold. This stochastic formulation mitigates the dead neuron problem by providing non-zero gradient updates even when neurons are below threshold.
\end{itemize}

Figure~\ref{fig:surrogates} visualizes the derivative profiles of these functions, illustrating how the scaling factor $\alpha$ modulates the gradient support and sharpness around the firing threshold ($U - U_{thr}$).

\subsubsection{Hardware Implications of Surrogate Selection}

The choice of surrogate function significantly impacts both training convergence and computational efficiency. Functions with sharp slopes (high $\alpha$ values) provide better gradient approximation but can lead to unstable updates. Conversely, smoother functions facilitate stable training but may reduce accuracy by underestimating gradients. 

Additionally, our empirical studies show that different datasets exhibit varying sensitivities to surrogate functions. For example, event-based vision datasets, such as \textit{DVS128-Gesture}, often benefit from functions that emphasize temporal dynamics, whereas static datasets may require more stable approximations. Thus, selecting an appropriate surrogate function in combination with optimal neuron model parameters can yield significant improvements in accuracy and efficiency. Characterizing this software-to-hardware link is a primary contribution of this work.

\subsection{Prior Work on SNN Accelerators and Activation Sparsity}
\label{sec:prior-snn-acc}

Recent efforts in neuromorphic hardware design have explored custom accelerator architectures for SNNs, with a growing emphasis on exploiting activation sparsity to improve energy efficiency and throughput. Prior work primarily falls into two categories: (1) hardware designs focused on architectural innovations for event-driven processing and (2) techniques to harness or enhance activation sparsity in the context of inference.

Sommer et al. \cite{sommer2022efficient} proposed an FPGA-based SNN accelerator that implemented sparse convolution via a cluster of 9 processing elements (PEs), each responsible for one filter coefficient addition per clock cycle. To support parallel neuron updates and reduce memory bottlenecks, the design introduced a technique called “memory interlacing,” enabling concurrent access to distributed membrane potential storage. The architecture supported output channel-wise parallelism by instantiating multiple PE clusters, each targeting a subset of output channels. Although this approach effectively leveraged spatial sparsity, it relied on parallel filter coefficient processing, increasing the complexity of memory scheduling and data movement. In contrast, our architecture avoids this dimension of parallelism to simplify memory management and achieve better control over neuron-specific sparsity dynamics, particularly under hyperparameter-optimized workloads.

On the ASIC front, Di Mauro et al. \cite{di2022sne} presented a Network-on-Chip (NoC)-based architecture for mapping full convolutional layers onto a custom chip. Their design featured PEs capable of processing 16 neurons in parallel and employed packet-switched routing to handle spike events between layers. This flexible routing strategy allowed the accelerator to handle irregular and asynchronous dataflows common in event-based processing. However, the inclusion of custom packet encoding/decoding logic and routing control introduced significant latency overhead and additional hardware complexity. While their NoC-based approach was effective in adapting to dynamic workloads, it did not directly address the impact of training-time hyperparameters on activation sparsity or inference efficiency.

Other recent studies have proposed hardware designs that selectively activate computation paths based on dynamic spike activity, demonstrating tangible gains in performance-per-watt by leveraging temporal sparsity \cite{yin2022sata, wang2023spiking, leigh2022selective}. More recent work has expanded this scope to include dual sparsity—activations and weights—by leveraging advanced pruning, quantization, and dataflow optimization techniques.

For example, \textit{FireFly-S} \cite{li2024firefly} presents a reconfigurable SNN accelerator that exploits both activation and weight sparsity via gradient rewiring and 4-bit quantization. It achieves over 85\% weight sparsity while maintaining accuracy across benchmarks like MNIST, DVS-Gesture, and CIFAR-10. Similarly, \textit{Prosperity} \cite{wei2025prosperity} introduces a method called \textit{Product Sparsity}, which enables computation reuse by detecting redundant inner product operations in sparse matrices. Applied to models like SpikeBERT, it reduces computation by up to 90\% and significantly lowers energy consumption. In another work, \textit{LoAS} \cite{yin2024loas} proposes a fully temporal-parallel dataflow architecture tailored for dual-sparse SNNs. It minimizes inter-timestep data movement and accelerates sequential processing, leading to substantial improvements in both speed and energy compared to earlier designs.

\noindent\textbf{The gap we address:} While these works demonstrate impressive efficiency gains, they largely treat activation sparsity as an emergent, fixed property of the trained model or a result of post-training compression (quantization/pruning). They do not systematically explore how \textit{the training process}---specifically the choice of surrogate gradients and neuron mechanics---can be tuned to fundamentally alter the workload's sparsity characteristics before it is implemented in hardware. Our work complements these architectural advances by establishing a methodology to optimize the \textit{input workload density}, providing a sparsity-first model generation strategy that can leverage any of the aforementioned accelerator backends.

\section{Workload Characterization Methodology} \label{sec:methodology}

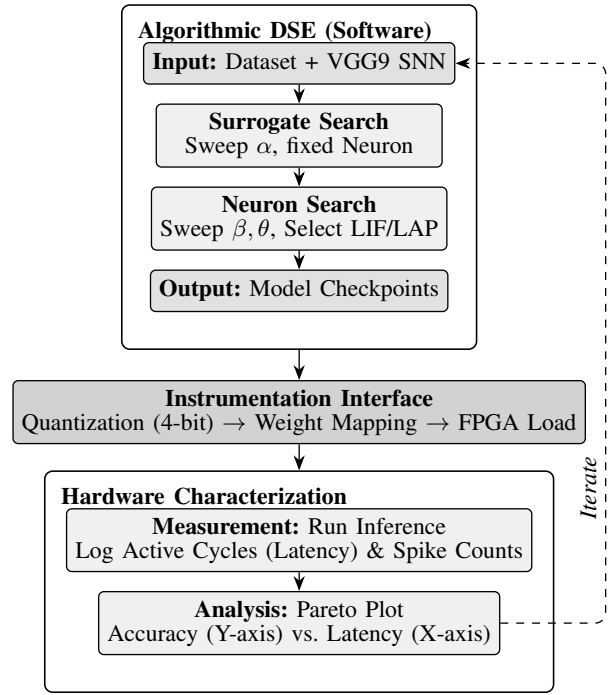
\begin{figure}[t]
    \centering
    \begin{tikzpicture}[
    font=\small,
    node distance=0.3cm and 0.4cm,
    >={Stealth[length=2mm, width=1.5mm]},
    box/.style={
        rectangle,
        rounded corners=2pt,
        draw=black,
        line width=0.5pt,
        fill=#1,
        minimum width=3.8cm,
        minimum height=0.7cm,
        align=center
    },
    smallbox/.style={
        rectangle,
        rounded corners=2pt,
        draw=black,
        line width=0.5pt,
        fill=#1,
        minimum width=2.8cm,
        minimum height=0.55cm,
        align=center
    },
    section/.style={
        rectangle,
        rounded corners=3pt,
        draw=black,
        line width=0.6pt,
        inner sep=6pt
    },
    sectionlabel/.style={
        font=\small\bfseries
    },
    arrow/.style={
        ->,
        line width=0.5pt
    },
    dashedarrow/.style={
        ->,
        line width=0.5pt,
        dashed
    }
]

\definecolor{lightgray1}{RGB}{240,240,240}
\definecolor{lightgray2}{RGB}{225,225,225}
\definecolor{lightgray3}{RGB}{210,210,210}

\node[smallbox=lightgray2] (input) {\textbf{Input:} Dataset +  VGG9 SNN};

\node[box=lightgray1, below=0.25cm of input] (surrogate) {
    \textbf{Surrogate Search}\\[-1pt]
    {\small Sweep $\alpha$, fixed Neuron}
};

\node[box=lightgray1, below=0.25cm of surrogate] (neuron) {
    \textbf{Neuron Search}\\[-1pt]
    {\small Sweep $\beta, \theta$, Select LIF/LAP}
};

\node[smallbox=lightgray2, below=0.25cm of neuron] (checkpoints) {\textbf{Output:} Model Checkpoints};

\begin{scope}[on background layer]
    \node[section, fill=white, fit=(input)(surrogate)(neuron)(checkpoints), 
          inner xsep=8pt, inner ysep=14pt, label={[sectionlabel, anchor=north west, xshift=3pt, yshift=-3pt]north west:Algorithmic DSE (Software)}] (software) {};
\end{scope}

\draw[arrow] (input) -- (surrogate);
\draw[arrow] (surrogate) -- (neuron);
\draw[arrow] (neuron) -- (checkpoints);

\node[box=lightgray3, below=0.4cm of software] (interface) {
    \textbf{Instrumentation Interface}\\[-1pt]
    {\small Quantization (4-bit) $\rightarrow$ Weight Mapping $\rightarrow$ FPGA Load}
};

\draw[arrow] (software.south) -- (interface.north);

\node[box=lightgray1, below=0.85cm of interface] (measurement) {
    \textbf{Measurement:} Run Inference\\[-1pt]
    {\small Log Active Cycles (Latency) \& Spike Counts}
};

\node[box=lightgray1, below=0.25cm of measurement] (analysis) {
    \textbf{Analysis:} Pareto Plot\\[-1pt]
    {\small Accuracy (Y-axis) vs.\ Latency (X-axis)}
};

\begin{scope}[on background layer]
    \node[section, fill=white, fit=(measurement)(analysis), 
          inner xsep=8pt, inner ysep=14pt, label={[sectionlabel, anchor=north west, xshift=3pt, yshift=-3pt]north west:Hardware Characterization}] (hardware) {};
\end{scope}

\draw[arrow] (interface.south) -- (hardware.north);
\draw[arrow] (measurement) -- (analysis);

\draw[dashedarrow, rounded corners=4pt] 
    (analysis.east) -- ++(1.4,0) |- (input.east);

\node[font=\small\itshape, rotate=90, anchor=south] 
    at ($(analysis.east)+(1.4,1.8)$) {Iterate};

\end{tikzpicture}  
    \caption{Overview of our design space exploration workflow.}
    \label{fig:dse}
\end{figure}

This work employs a hardware-in-the-loop characterization methodology to quantify the sensitivity of SNN inference performance to training-time hyperparameters (the RTL can be found at \url{https://zenodo.org/records/18893738}). The methodology bridges the abstraction gap between algorithmic choices (surrogates, neuron models) and physical execution metrics (latency, active cycles). We utilize a two-phase design space exploration (DSE) strategy validated on a cycle-accurate FPGA instrumentation platform. This section details the hardware measurement setup, the hyperparameter search space, and the sparsity quantification metrics.

\subsection{Hardware Instrumentation Platform Architecture} \label{sec:hardware}

To obtain precise, cycle-accurate measurements of inference latency and dynamic power, we employ a custom \textbf{Parametric Sparsity-Aware Convolution Engine} \cite{aliyev2024pulse}. Unlike general-purpose baselines, this platform serves as a specialized instrumentation tool designed to model the fine-grained, data-dependent sparsity characteristics of neuromorphic workloads. The architecture employs an event-driven processing paradigm where computational latency is directly proportional to input activation density.

\begin{figure*}[t]
    \centering
    \resizebox{0.8\textwidth}{!}{\begin{tikzpicture}[
    font=\footnotesize,
    node distance=0.15cm and 0.25cm,
    >={Stealth[length=1.5mm, width=1mm]},
    box/.style={
        rectangle,
        draw=black,
        line width=0.4pt,
        fill=#1,
        minimum height=0.55cm,
        align=center,
        font=\scriptsize
    },
    tallbox/.style={
        rectangle,
        draw=black,
        line width=0.4pt,
        fill=#1,
        minimum width=0.55cm,
        minimum height=1.3cm,
        align=center,
        font=\scriptsize
    },
    dashedbox/.style={
        rectangle,
        draw=black,
        line width=0.4pt,
        dashed,
        fill=white,
        minimum height=0.55cm,
        align=center
    },
    section/.style={
        rectangle,
        draw=black,
        line width=0.5pt,
        dashed,
        inner sep=3pt
    },
    arrow/.style={
        ->,
        line width=0.4pt
    },
    doublearrow/.style={
        <->,
        line width=0.4pt
    }
]

\definecolor{lutgreen}{RGB}{198,224,180}
\definecolor{ffblue}{RGB}{180,210,230}
\definecolor{ramgray}{RGB}{242,242,242}

\node[dashedbox, minimum width=1cm, minimum height=1.6cm] (spikein) {};
\foreach \row in {0,1,2,3,4} {
    \pgfmathsetmacro{\ypos}{0.55 - \row*0.25}
    \draw[line width=0.25pt, gray] ([yshift=\ypos cm]spikein.center) +(-0.4,0) -- +(0.4,0);
}
\foreach \x/\y in {-0.3/0.55, -0.15/0.55, 0.1/0.55, 0.25/0.55,
                   -0.25/0.3, -0.05/0.3, 0.15/0.3, 0.35/0.3,
                   -0.35/0.05, -0.1/0.05, 0.2/0.05,
                   -0.2/-0.2, 0.05/-0.2, 0.3/-0.2,
                   -0.3/-0.45, -0.05/-0.45, 0.15/-0.45, 0.25/-0.45} {
    \draw[line width=0.4pt, blue!70!black] ([xshift=\x cm, yshift=\y cm]spikein.center) -- ++(0,0.15);
}
\node[font=\scriptsize, below=0.02cm of spikein, text=black] {\textit{Spike trains}};

\node[tallbox=ffblue, right=0.35cm of spikein] (controller) {\rotatebox{90}{Controller}};

\draw[arrow] ([yshift=0.3cm]spikein.east) -- ([yshift=0.3cm]controller.west);
\draw[arrow] (spikein.east) -- (controller.west);
\draw[arrow] ([yshift=-0.3cm]spikein.east) -- ([yshift=-0.3cm]controller.west);

\node[box=lutgreen, right=0.2cm of controller, minimum width=0.75cm, minimum height=0.5cm, yshift=0.35cm] (addrgen) {Address\\[-3pt]gen};
\node[box=lutgreen, below=0.15cm of addrgen, minimum width=0.75cm, minimum height=0.5cm] (penc) {PENC};

\node[tallbox=ffblue, right=0.2cm of addrgen, yshift=-0.35cm] (spikeevents) {\rotatebox{90}{Spike events}};

\begin{scope}[on background layer]
    \node[section, fit=(controller)(addrgen)(penc)(spikeevents), inner xsep=4pt, inner ysep=6pt] (ecu) {};
\end{scope}
\node[font=\scriptsize, anchor=south] at (ecu.north) {Event control unit};

\draw[arrow] ([yshift=0.2cm]controller.east) -- ++(0.05,0) |- (addrgen.west);
\draw[arrow] ([yshift=-0.2cm]controller.east) -- ++(0.05,0) |- (penc.west);
\draw[arrow] (addrgen.east) -- ++(0.05,0) |- ([yshift=0.2cm]spikeevents.west);
\draw[arrow] (penc.east) -- ++(0.05,0) |- ([yshift=-0.2cm]spikeevents.west);

\node[anchor=west, font=\scriptsize] at ([xshift=0.3cm, yshift=0.1cm]ecu.north east) {
    \tikz[baseline=-0.5ex]{
        \fill[lutgreen] (0,0) rectangle (0.3,0.22); 
        \draw[line width=0.3pt] (0,0) rectangle (0.3,0.22);
        \node[right, font=\scriptsize] at (0.35,0.11) {LUT};
        \fill[ffblue] (1.05,0) rectangle (1.35,0.22); 
        \draw[line width=0.3pt] (1.05,0) rectangle (1.35,0.22);
        \node[right, font=\scriptsize] at (1.4,0.11) {FF};
        \fill[ramgray] (2.0,0) rectangle (2.3,0.22); 
        \draw[line width=0.3pt, dashed] (2.0,0) rectangle (2.3,0.22);
        \node[right, font=\scriptsize] at (2.35,0.11) {RAM};
    }
};

\node[box=ffblue, right=0.45cm of spikeevents, minimum width=1.2cm] (accum) {Accumulator};
\node[box=lutgreen, below=0.15cm of accum, minimum width=1.2cm] (filter) {Filter weights};
\node[box=ffblue, below=0.15cm of filter, minimum width=1.2cm] (activ) {Activation};

\begin{scope}[on background layer]
    \node[section, fit=(accum)(filter)(activ), inner xsep=4pt, inner ysep=6pt] (neuralcores) {};
\end{scope}
\node[font=\scriptsize, anchor=south] at (neuralcores.north) {Neural cores};

\draw[arrow] (spikeevents.east) -- (accum.west);

\draw[doublearrow] (accum.south) -- (filter.north);
\draw[doublearrow] (filter.south) -- (activ.north);

\node[dashedbox, right=0.4cm of filter, minimum width=1.2cm, minimum height=1.55cm, fill=ramgray, font=\scriptsize, align=center] (memory) {Memory\\(membrane\\potential\\and weights)};

\draw[doublearrow] (accum.east) -- ++(0.12,0) |- ([yshift=0.4cm]memory.west);
\draw[doublearrow] (filter.east) -- (memory.west);
\draw[doublearrow] (activ.east) -- ++(0.12,0) |- ([yshift=-0.4cm]memory.west);

\node[dashedbox, right=0.4cm of memory, minimum width=1cm, minimum height=1.6cm] (spikeout) {};
\foreach \row in {0,1,2,3,4} {
    \pgfmathsetmacro{\ypos}{0.55 - \row*0.25}
    \draw[line width=0.25pt, gray] ([yshift=\ypos cm]spikeout.center) +(-0.4,0) -- +(0.4,0);
}
\foreach \x/\y in {-0.2/0.55, 0.15/0.55, 0.35/0.55,
                   -0.35/0.3, 0.0/0.3, 0.25/0.3,
                   -0.25/0.05, 0.1/0.05, 0.3/0.05,
                   -0.1/-0.2, 0.2/-0.2,
                   -0.3/-0.45, 0.05/-0.45, 0.35/-0.45} {
    \draw[line width=0.4pt, blue!70!black] ([xshift=\x cm, yshift=\y cm]spikeout.center) -- ++(0,0.15);
}
\node[font=\scriptsize, below=0.02cm of spikeout, text=black] {\textit{Spike trains}};

\draw[arrow] (memory.east) -- (spikeout.west);

\end{tikzpicture}}  %
    \caption{Dataflow of the hardware instrumentation platform (Parametric Sparsity-Aware Convolution Engine). The design features an Event Control Unit (ECU) that dynamically schedules only active neurons, and Neural Cores (NCs) with configurable accumulation logic. This event-driven architecture allows for precise isolation of latency reductions caused by algorithmic sparsity tuning.}
    \label{fig:architecture}
\end{figure*}
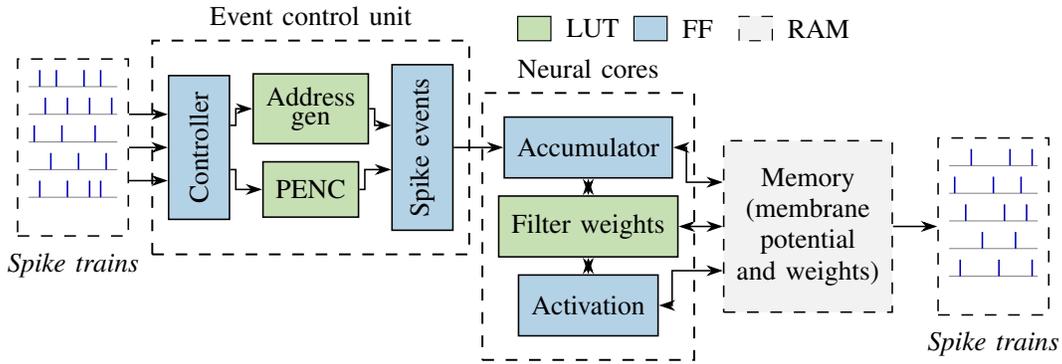

As illustrated in Figure~\ref{fig:architecture}, the architecture is decoupled into an Event Control Unit (ECU) and parallel Neural Cores (NCs) connected through a programmable, event-based pipeline. The ECU implements a priority encoder to detect active spikes and dispatch target addresses, effectively skipping zero-valued activations in hardware. The NCs support configurable neuron models (LIF, LAP) via modular activation logic, allowing us to vary the arithmetic intensity of the neuron update without altering the underlying dataflow. This setup allows us to isolate the impact of \textit{workload density} on latency, providing a ground-truth measurement of how hyperparameter choices translate to hardware performance.

\subsubsection{Event-driven execution model} The core mechanism linking algorithmic sparsity to hardware latency is the ECU. Unlike systolic arrays that process dense matrices, the ECU utilizes a Priority Encoder (PENC) to dynamically scan input spike trains.

\noindent\textbf{Sparsity-to-latency translation:} The PENC identifies non-zero indices in a single cycle and filters out inactive neurons. This allows the hardware to ``skip" cycles proportional to the sparsity induced by the surrogate gradient function.

\noindent\textbf{Address generation:} An Address Generation Unit (AGU) computes target neuron indices only for active spikes. This creates a variable-latency datapath where the execution time of a layer is defined by: 

\begin{equation} T_{layer} \approx \frac{1}{P} \times (C_{ovHD} + N_{active} \times T_{accum}) \label{eq:5} \end{equation}

\noindent where $C_{ovHD}$ is the constant control overhead, $N_{active}$ is the number of non-zero events (determined by the surrogate), $T_{accum}$ is the accumulation time, and $P$ is the parallelization factor. This mechanism ensures that our latency measurements are sensitive to even minor changes in training-time sparsity.

\subsubsection{Configurable neural cores} To isolate the performance impact of neuron model selection (e.g., LIF vs. LAP), the neural cores (NCs) feature a modular Activation Unit.

\noindent\textbf{Model agnostic datapath:} The accumulation logic is decoupled from the neuron state update rules. This allows us to instantiate different neuron models (Eq.~\ref{eq:lif} and~\ref{eq:lap}) within the same physical pipeline, ensuring that reported latency differences are due to the \textit{model's arithmetic intensity} and \textit{temporal dynamics}, not architectural disparities.

\noindent\textbf{Precision and storage:} Filter weights are stored in flip-flops (FFs) for immediate access, while membrane potentials utilize Block RAMs (BRAMs). Weights are quantized to 4-bit integers, and membrane potentials use fixed-point representation. To minimize overhead, the activation unit employs an optimized comparator that inspects the most significant bits (MSB) for threshold crossing, balancing decision accuracy with logic depth.

\noindent\textbf{Memory bandwidth:} The event-driven execution model minimizes bandwidth pressure by only reading/writing membrane potentials for active neurons receiving spike events, and only fetching weights when the corresponding input channel is active. This sparsity-proportional memory access pattern ensures that the bandwidth requirements scale with activity density rather than the network size, reinforcing the efficiency gains from hyperparameter-induced sparsity.

\subsubsection{Workload distribution} The architecture employs an output channel-wise parallelization strategy. Each NC processes a spatial chunk of the Output Feature Map (OFM), eliminating data hazards. The effective computational workload $W$ for a convolutional layer is modeled as: 

\begin{equation} W_{CONV} = F \times C_{out} \times \sum_{i=1}^{M} S_i \label{eq:6} \end{equation}

where $F$ is the kernel size, $C_{out}$ is the number of output channels, and $S_i$ is the spike count of input channel $i$. 

Throughout this work, ``latency" refers to the measured active clock cycles on our cycle-accurate FPGA platform, converted to wall-clock time at the operating frequency. The analytical model in Eq.~\ref{eq:5} accurately tracks these physical measurements across all surrogate and neuron model sweeps presented in Section~\ref{sec:results}, confirming that inference time scales linearly with activity density as predicted by the event-driven execution model.

\subsubsection{Generalizability} While we validate our work on a specific FPGA platform, our methodology is architecture-agnostic. The core insight---that latency tracks activity density via event-skipping (Eq.~\ref{eq:5}) and the workload model (Eq.~\ref{eq:6})---generalizes to any event-driven accelerator that conditionally schedules computation on spike events. 

\subsection{Design Space Exploration Framework}

We implement an automated DSE pipeline using the Optuna framework \cite{akiba2019optuna} to traverse the hyperparameter space. The exploration is divided into two phases to decouple the effects of gradient approximation from neuron dynamics.

\paragraph*{Phase 1: Surrogate gradient sensitivity analysis}
We first characterize the impact of gradient sharpness on the resulting model sparsity. We sweep the slope parameter $\alpha \in [1,48]$ across four surrogate functions (FS, ATAN, SRE, SSO). For each candidate, we conduct 40 independent trials (200 epochs each) to identify the configurations that maximize accuracy. The top-performing candidates are then profiled on the hardware platform to map the relationship between gradient slope $\alpha$ and inference latency.

\noindent\textbf{DSE overhead:} This exploration is performed \textit{offline} as a one-time profiling step; deployment uses a single selected configuration with no runtime overhead. The total training cost is approximately $40 \times 200 = 8{,}000$ epochs per surrogate function. However, as shown in Fig.~\ref{fig:surrogate_lat}, evaluating only the top-2 slope configurations per surrogate captures the majority of Pareto-relevant insights, reducing the effective hardware profiling to 8 configurations per dataset. This demonstrates that a lightweight subset of the full DSE suffices to identify near-optimal accuracy-latency trade-offs.

\paragraph*{Phase 2: Neuron model exploration}
In the second phase, we evaluate the hardware cost-benefit trade-off of neuron complexity. We expand the search space to include both LIF and LAP models. We discretize the leak factor $\beta \in [0.1, 1.0]$ and firing threshold $\theta \in [0.1, 2.0]$ with a step size of 0.2. This discretization allows sufficient resolution to capture performance variation while controlling the total number of experiments. The optimal configuration for each dataset is determined based on the joint accuracy and sparsity profile, with a follow-up hardware evaluation to validate its effect on inference efficiency. This phase specifically investigates whether the increased arithmetic complexity of the LAP model is amortized by the potential reduction in total spike events.

\subsection{Network Architecture} \label{sec:network}

We utilize a VGG9-based \cite{vgg9} convolutional SNN adapted for event-based input streams. The topology is defined as:
\[
\begin{aligned}
64\text{C}3\text{-}28\text{C}3\text{-MP}2\text{-}48\text{C}3\text{-}54\text{C}3\text{-MP}2 \\ \text{-}120\text{C}3\text{-}126\text{C}3\text{-}140\text{C}3\text{-MP}2\text{-}216\text{-}200
\end{aligned}
\]
where $X$C$Y$ denotes $X$ filters of size $Y \times Y$ and MP$Z$ denotes a $Z \times Z$ max-pooling layer. We apply 4-bit integer quantization to all weights to align with the constraints of edge neuromorphic hardware. Training is performed using \texttt{snnTorch} with cosine annealing schedules \cite{loshchilov2016sgdr}.

\subsection{Workload Density Modeling}

To quantify the ``workload" processed by the hardware, we measure the \textbf{Activity Density $\mathcal{A}$}, defined as the average spike rate per neuron per timestep. This metric is the inverse of sparsity ($1 - \text{Sparsity}$) and is calculated as:
\[
\mathcal{A} = \frac{1}{T \cdot N} \sum_{t=1}^{T} \sum_{i=1}^{L} S_{i}[t]
\]
where $T$ is the number of timesteps, $N$ is the total neuron count across all layers, $L$ is the number of layers, and $S_i[t]$ is the spike count in layer $i$ at time $t$. By correlating $\mathcal{A}$ with the measured hardware latency, we can rigorously evaluate how effectively different surrogate functions suppress unnecessary computation.

\subsection{Evaluation Setup}

Following the algorithmic exploration, top-performing model configurations are synthesized and deployed on the FPGA instrumentation platform, implemented in SystemVerilog and mapped to a Xilinx Kintex Ultrascale+ platform. Each configuration is evaluated for classification accuracy, average inference latency, and power, enabling us to assess the end-to-end impact of hyperparameter choices on system-level performance.

\begin{figure*}[t]
    \centering
    \begin{subfigure}{.32\linewidth}
        \centering
        \includegraphics[width=\linewidth]{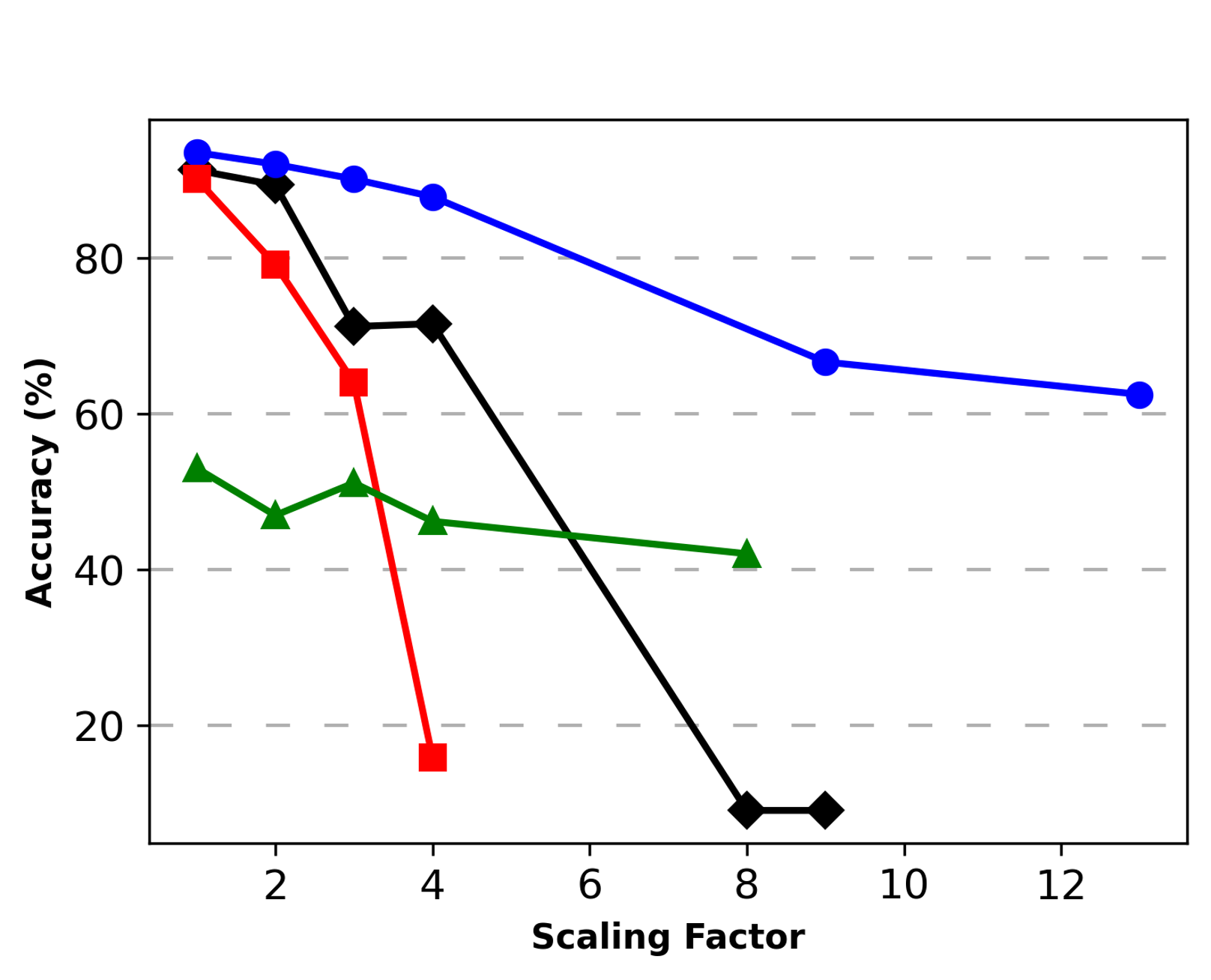}
        \caption{DVS-Gesture}
        \label{fig:gest_acc}
    \end{subfigure}%
    \begin{subfigure}{.32\linewidth} 
        \centering
        \includegraphics[width=\linewidth]{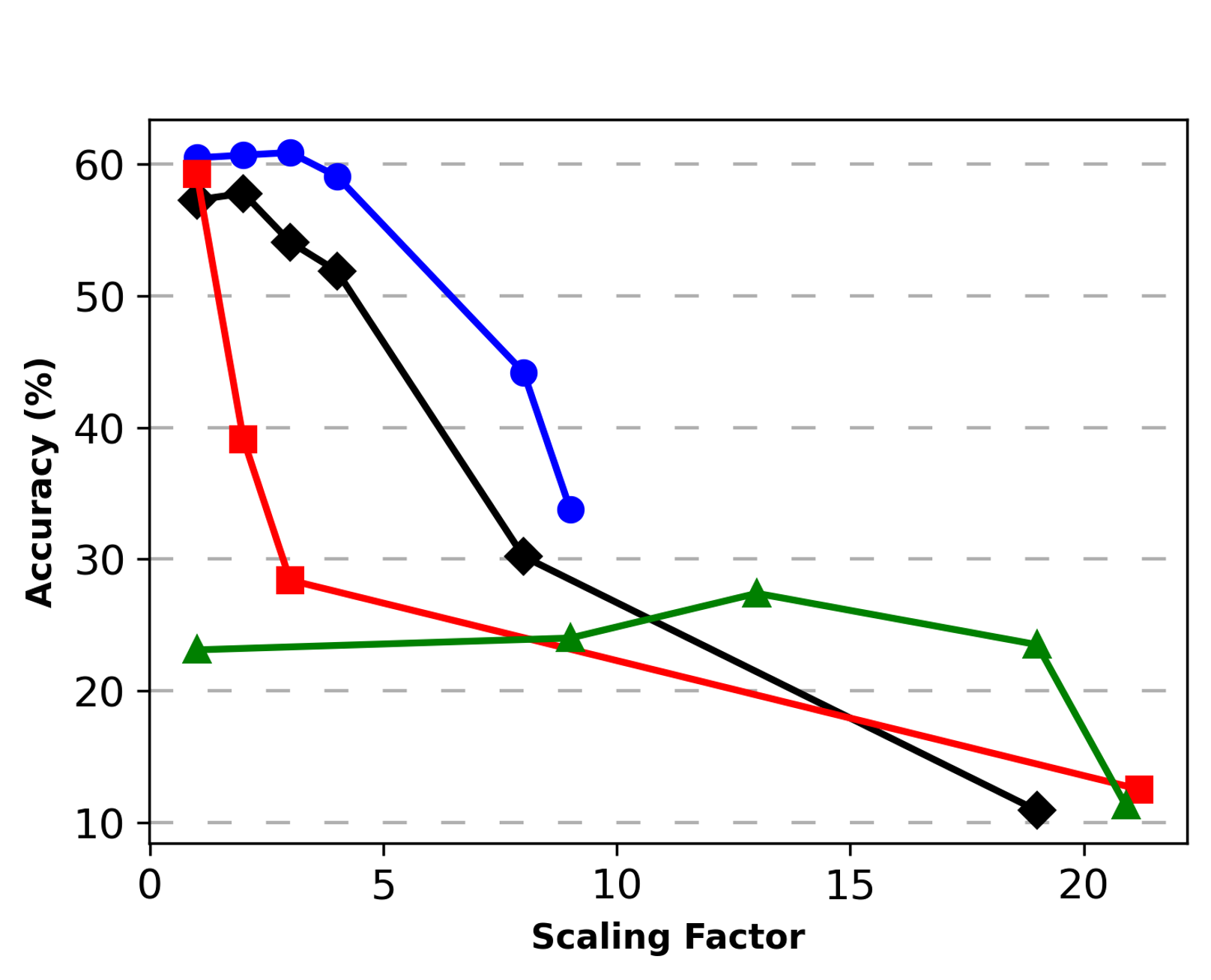} 
        \caption{DVS-CIFAR10}
        \label{fig:c100_acc}
    \end{subfigure}
    \begin{subfigure}{.32\linewidth} 
        \centering
        \includegraphics[width=\linewidth]{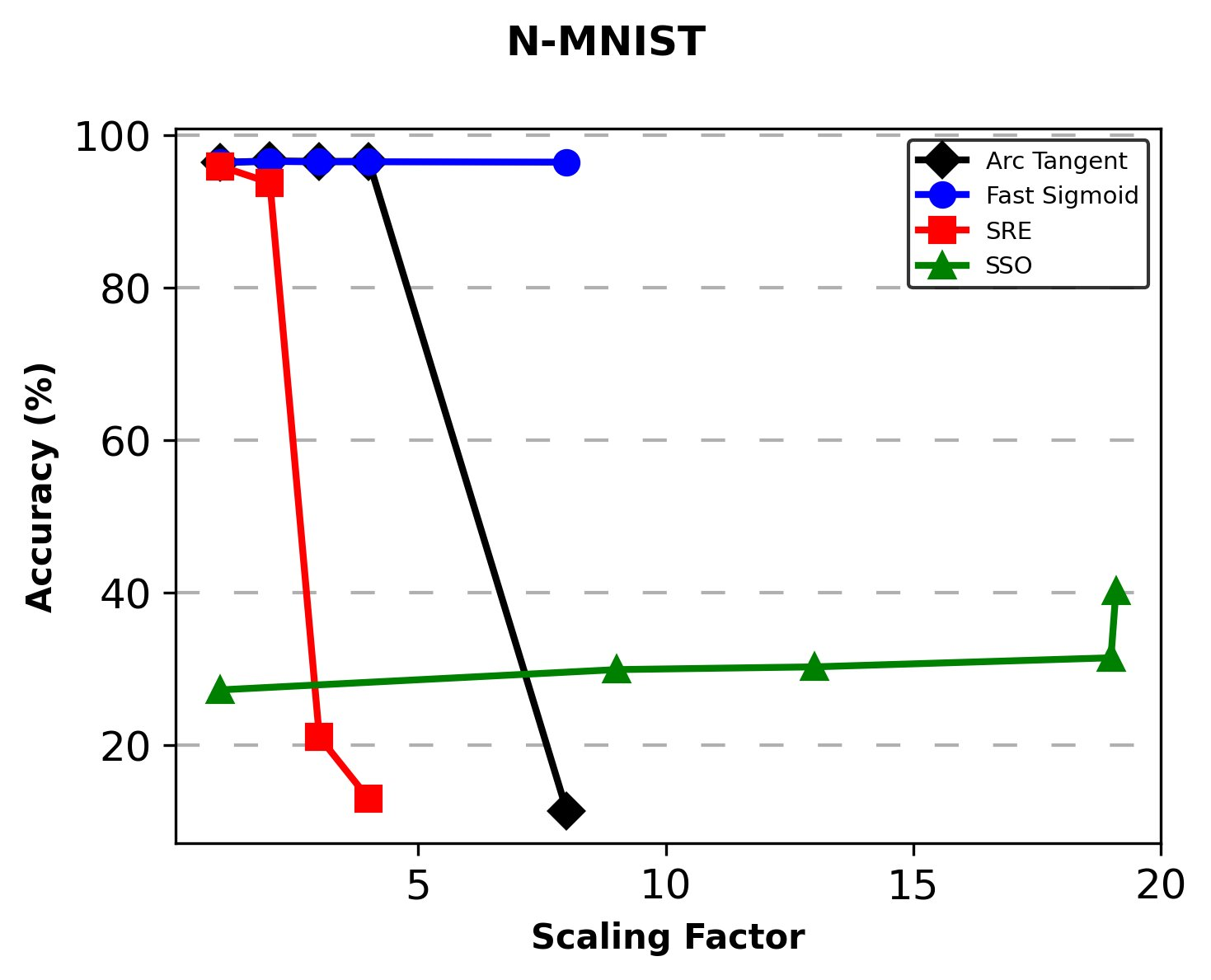} 
        \caption{N-MNIST}
        \label{fig:nmnist_acc}
    \end{subfigure}
    
    \caption{\textbf{Accuracy trends for surrogate gradient functions across three datasets.} Each plot shows classification accuracy as a function of the derivative scaling factor (slope) $\alpha$ for four surrogate functions: \texttt{Fast Sigmoid}, \texttt{Arctangent}, \texttt{SRE}, and \texttt{SSO}. For each surrogate, representative top-performing $\alpha$ values were selected from 40 Optuna trials per candidate configuration (each trained for up to 200 epochs with early stopping).}
    \label{fig:surrogate_acc}
\end{figure*}

\section{Characterization Results and Analysis}
\label{sec:results}

This section characterizes the sensitivity of hardware inference metrics to training-time hyperparameter selection. We focus on quantifying the trade-offs between classification accuracy and inference latency across three event-based benchmarks: \textit{DVS-Gesture, DVS-CIFAR10,} and \textit{N-MNIST}. Our analysis isolates the impact of the two primary variables: the surrogate gradient function (which dictates learning dynamics) and the neuron model (which dictates temporal dynamics).

\subsection{Sensitivity of Accuracy to Gradient Approximation}
\label{sec:results-surrogate-accuracy}

Fig.~\ref{fig:surrogate_acc} plots the classification accuracy sensitivity to the surrogate derivative scaling factor $\alpha$ of four surrogate gradient functions: \texttt{Fast Sigmoid (FS)}, \texttt{Arctangent (ATAN)}, \texttt{Spike Rate Escape (SRE)}, and \texttt{Stochastic Spike Operator (SSO)}. The choice of surrogate fundamentally alters the stability profile of the training process.

\noindent\textbf{High-performance stability (FS):} The \texttt{Fast Sigmoid (FS)} function demonstrates superior robustness. It maintains peak accuracy ($>$ 90\% on N-MNIST/DVS-Gesture) across the widest range of slopes, establishing it as a safe default for maximizing classification performance in general-purpose SNNs.

\noindent\textbf{Volatile degradation (SRE):} The \texttt{Spike Rate Escape (SRE)} exhibits a cliff-like degradation profile. This is most visible in N-MNIST (Fig.~\ref{fig:nmnist_acc}) and DVS-CIFAR10 (Fig.~\ref{fig:c100_acc}), where \texttt{SRE} accuracy collapses from $>$90\% to $<$20\% with only minor increases in slope. This exponential gradient formulation is highly aggressive, suppressing spiking activity so rapidly that it risks vanishing gradients if the slope is not carefully bounded.

\noindent\textbf{Low-fidelity stability (SSO):} While the \texttt{Stochastic Spike Operator (SSO)} appears stable (exhibiting a flat slope response), it consistently fails to converge to competitive accuracy levels (stalling at $<$50\% on DVS-Gesture and $<$30\% on DVS-CIFAR). Unlike \texttt{FS}, which is stable and accurate, \texttt{SSO} achieves stability only by failing to capture the fine-grained temporal dynamics required for these tasks. Consequently, we treat it as a sub-optimal candidate for high-performance inference.

\begin{figure*}[t]
    \centering
    \begin{subfigure}{.32\linewidth}
        \centering
        \includegraphics[width=\linewidth]{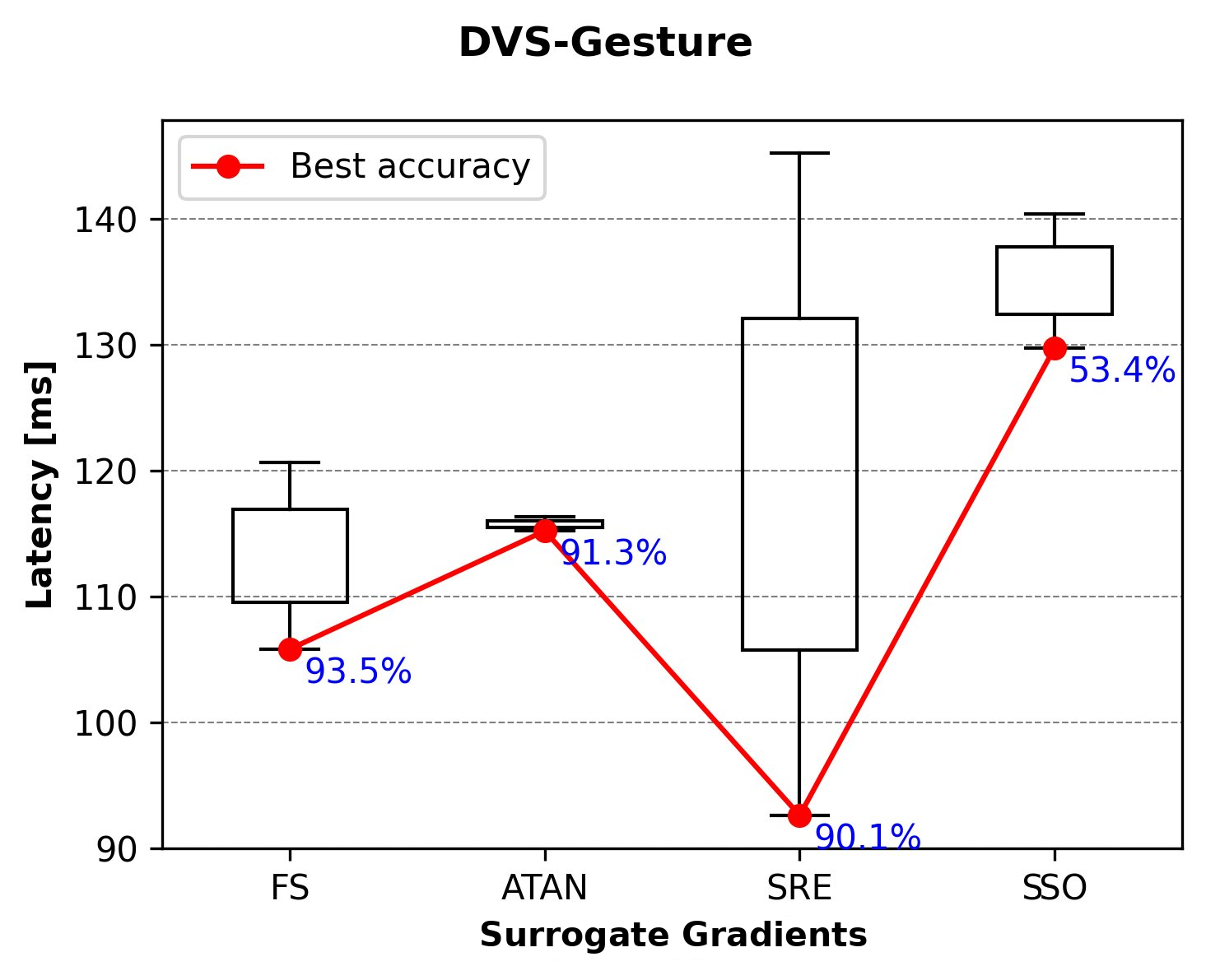}
        \caption{DVS-Gesture}
        \label{fig:gest_lat}
    \end{subfigure}%
    \begin{subfigure}{.32\linewidth} 
        \centering
        \includegraphics[width=\linewidth]{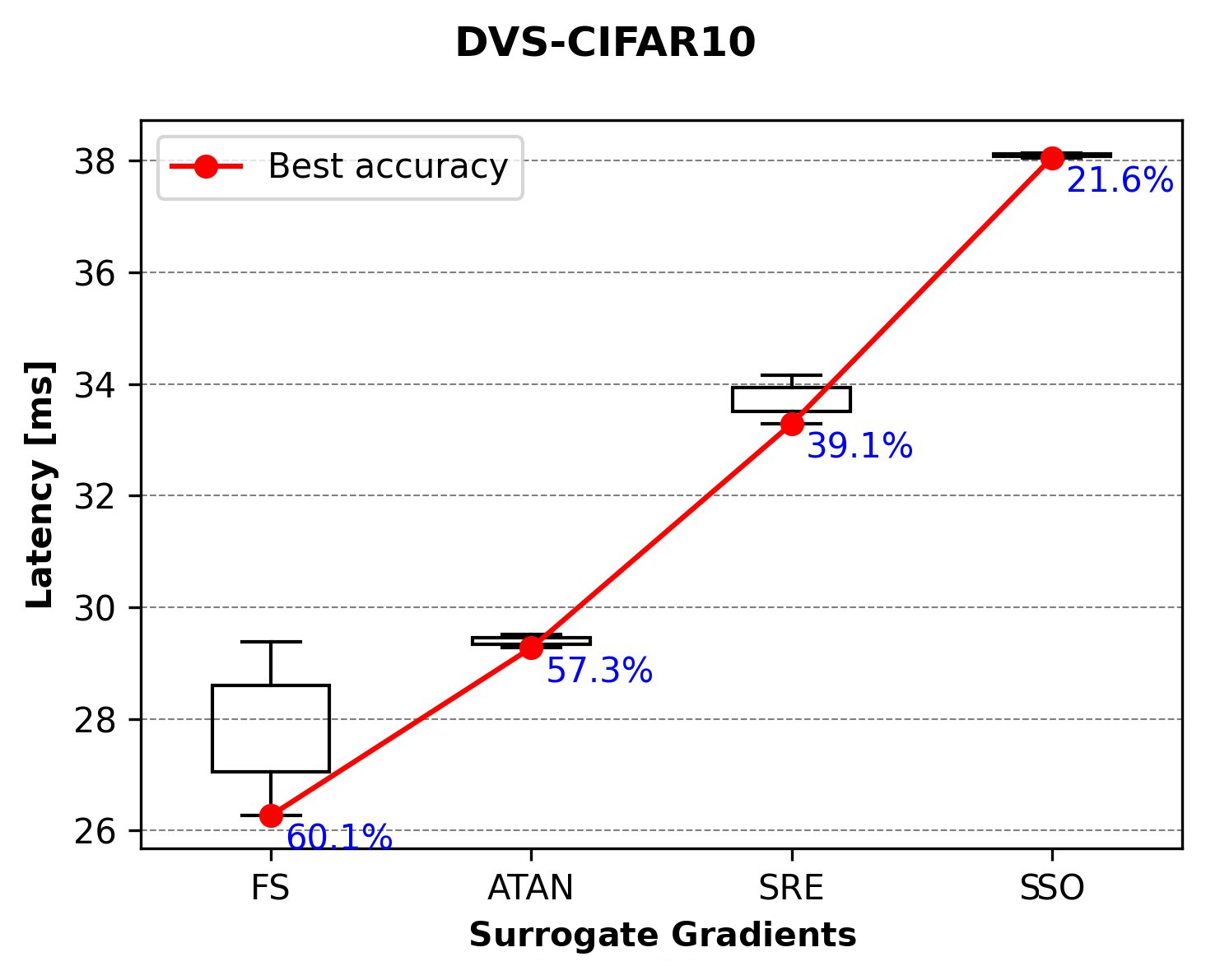} 
        \caption{DVS-CIFAR10}
        \label{fig:c100_lat}
    \end{subfigure}
    \begin{subfigure}{.32\linewidth} 
        \centering
        \includegraphics[width=\linewidth]{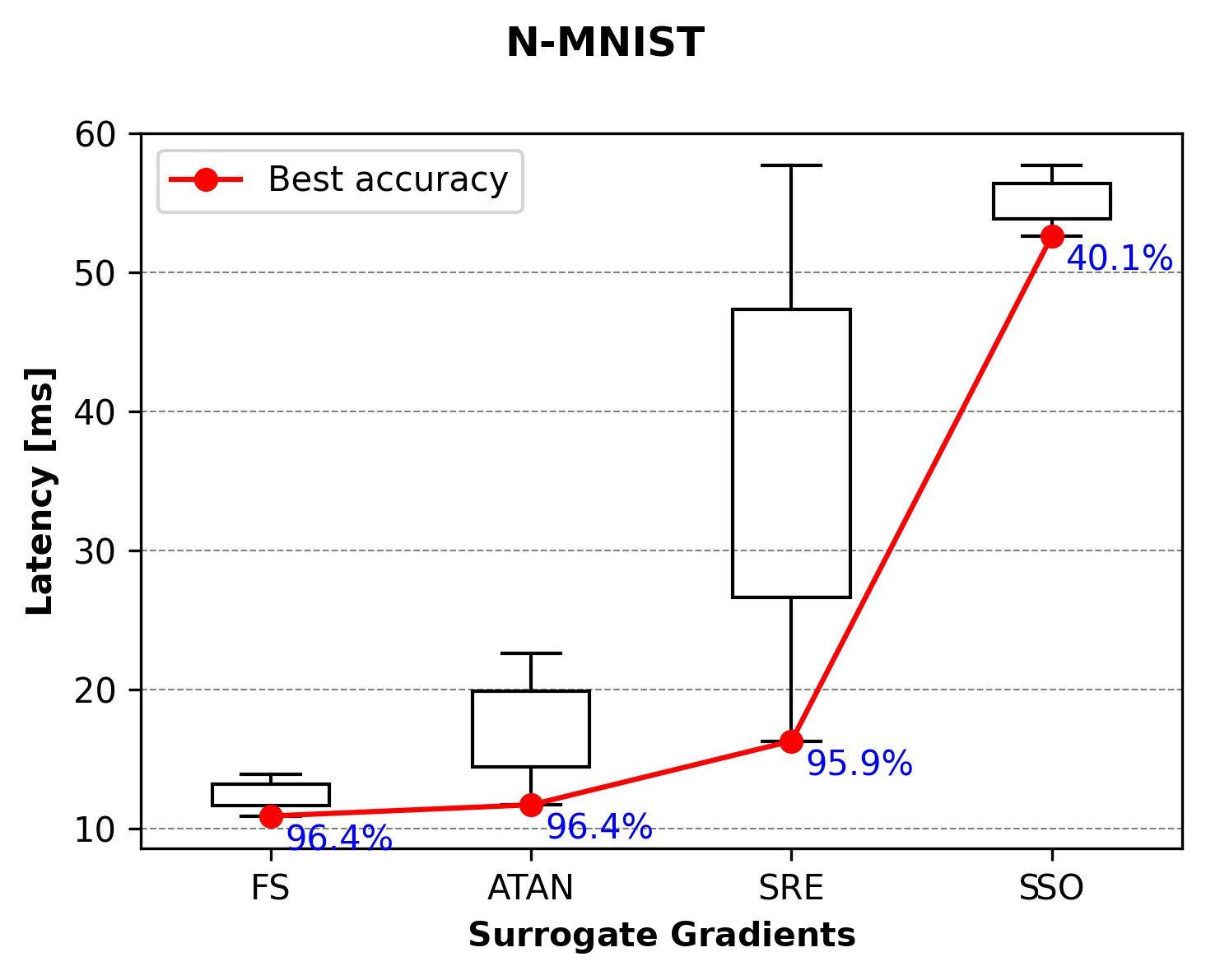} 
        \caption{N-MNIST}
        \label{fig:nmnist_lat}
    \end{subfigure}

    \caption{\textbf{Hardware inference latency across surrogate functions for three datasets.} Each box represents the latency range for the top two slope configurations (best accuracy) of a surrogate. The red line connects the latency of the best-accuracy slope per surrogate. Accuracy values (blue) are annotated for the best configuration. Lower latency combined with higher accuracy indicates better efficiency.}
    \label{fig:surrogate_lat}
\end{figure*}

\subsection{Latency Characterization of Surrogate Functions}
\label{sec:results-surrogate-latency}

To isolate the algorithmic impact of surrogate functions, the latency values in this section are measured on a baseline hardware configuration with minimal parallelism (P = 1). System-level optimizations and resource scaling are applied in Section~\ref{sec:comparison-prior} to demonstrate peak realizable performance.

Figure~\ref{fig:surrogate_lat} correlates the chosen surrogate function with the measured inference latency on our hardware instrumentation platform. For each surrogate, we evaluate the top two slope configurations obtained from the accuracy-based exploration and represent their corresponding latency range using a box plot. The red line traces the latency associated with the best-accuracy slope setting for each surrogate. Percentage labels indicate the accuracy achieved at that slope setting. This analysis reveals a critical disconnect between software accuracy and hardware efficiency.

\noindent\textbf{The sparsity-latency mechanism:} Lower latency is driven by the Priority Encoder (PENC) skipping zero-activations. The results indicate that surrogates with sharp, exponential tails (like \texttt{SRE}) act as aggressive noise gates, suppressing spikes near the threshold more effectively than smooth sigmoidal functions

Looking at the DVS-Gesture dataset as a case study, the aggressive suppression behavior of \texttt{SRE}, which was a liability for accuracy stability (Section~\ref{sec:results-surrogate-accuracy}), becomes a massive asset for latency. By operating with a low scaling factor, \texttt{SRE} achieves the global minimum latency of $\sim$92 ms (Fig.~\ref{fig:gest_lat}), providing a \textbf{12.2\% speedup} over the accuracy-optimal \texttt{FS} baseline ($\sim$105 ms). This identifies \texttt{SRE} as a Pareto-optimal choice for latency-critical applications where a 3.4\% accuracy trade-off is acceptable. For edge-deployed neuromorphic systems processing high-frequency event streams, a 12.2\% latency reduction can represent the difference between meeting or missing real-time constraints. Importantly, since surrogate selection is an offline training decision, this efficiency gain incurs \textit{zero runtime cost}. \texttt{SRE} is thus particularly valuable when latency is paramount and the workload tolerates aggressive spike suppression.

\noindent\textbf{Dataset sensitivity:} This behavior is workload-dependent. On DVS-CIFAR10 (Fig.~\ref{fig:c100_lat}) and N-MNIST (Fig.~\ref{fig:nmnist_lat}), \texttt{SRE} fails to find a sweet spot. It yields higher latency than \texttt{FS} ($\sim$33 ms vs. $\sim$26 ms for DVS-CIFAR10) while simultaneously suffering from lower accuracy. This suggests that for object recognition tasks (unlike gesture recognition), the aggressive pruning of \texttt{SRE} removes information critical for classification before it removes redundancy, making \texttt{FS} the dominant choice for both metrics.

\begin{figure*}[t!]
    \centering
    \begin{subfigure}{0.32\linewidth} 
        \centering
        \includegraphics[width=\linewidth]{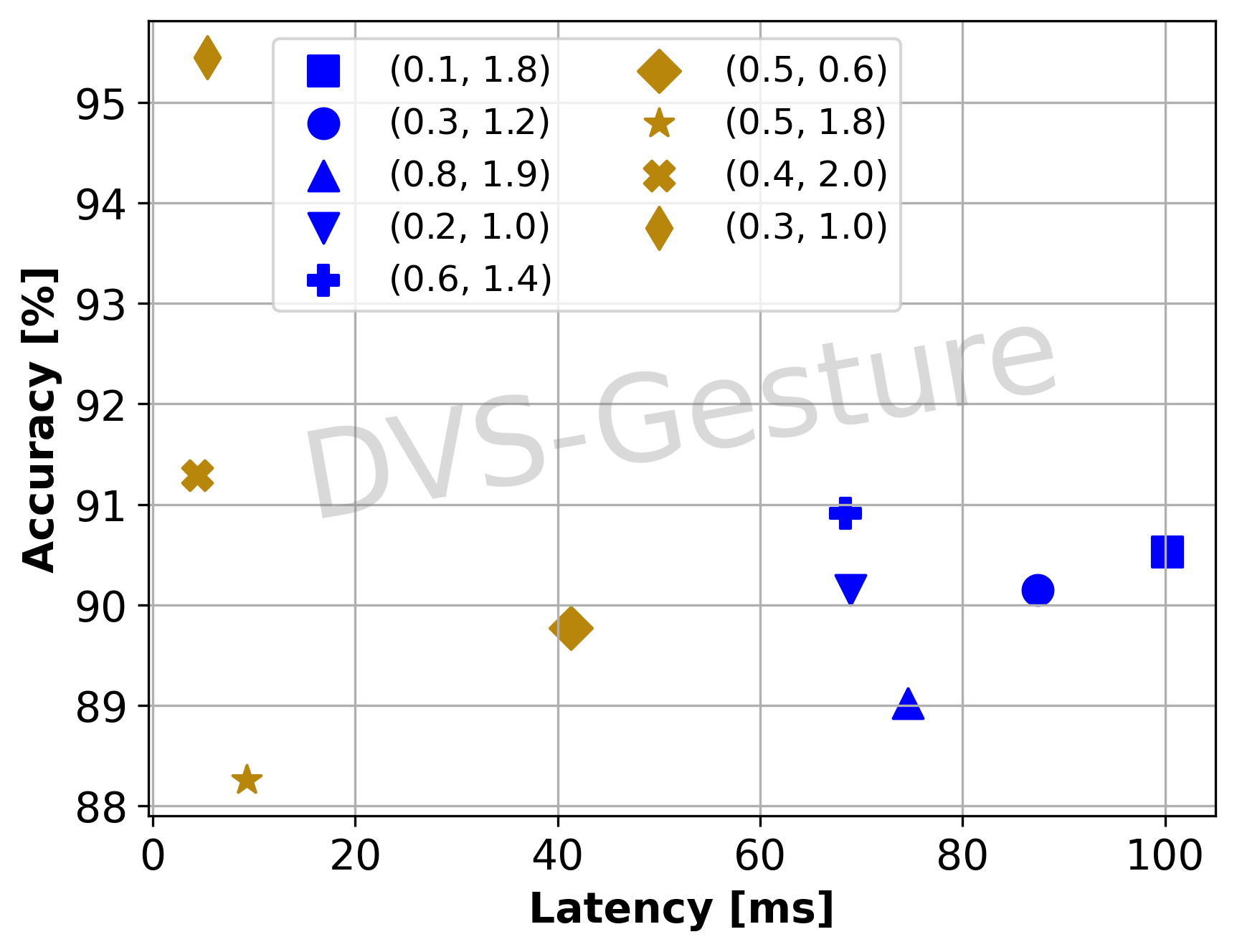}
        \caption{}
        \label{fig:gest_pareto}
    \end{subfigure}
    \begin{subfigure}{0.32\linewidth} 
        \centering
        \includegraphics[width=\linewidth]{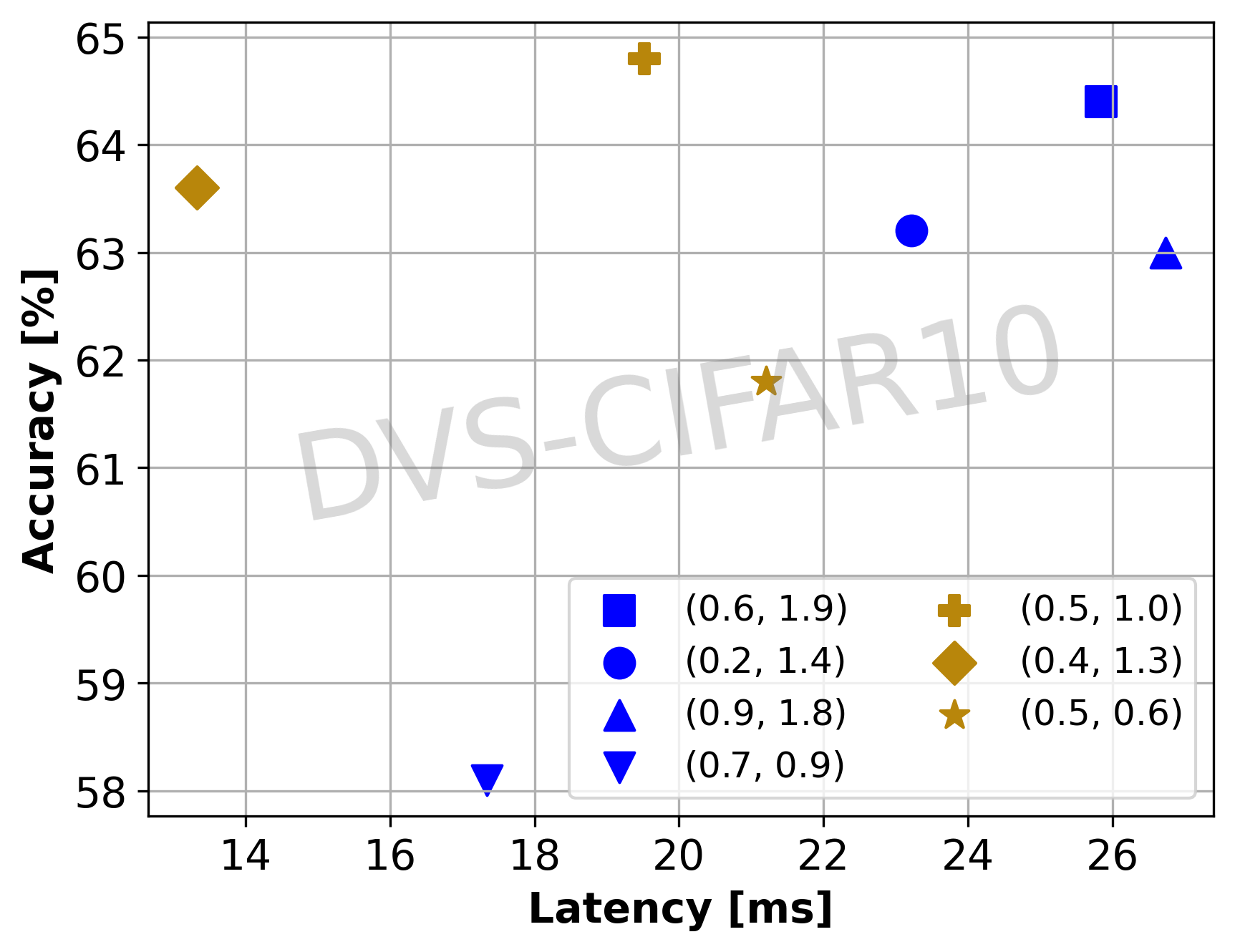} 
        \caption{}
        \label{fig:cifar_pareto}
    \end{subfigure}
    \begin{subfigure}{0.32\linewidth} 
        \centering
        \includegraphics[width=\linewidth]{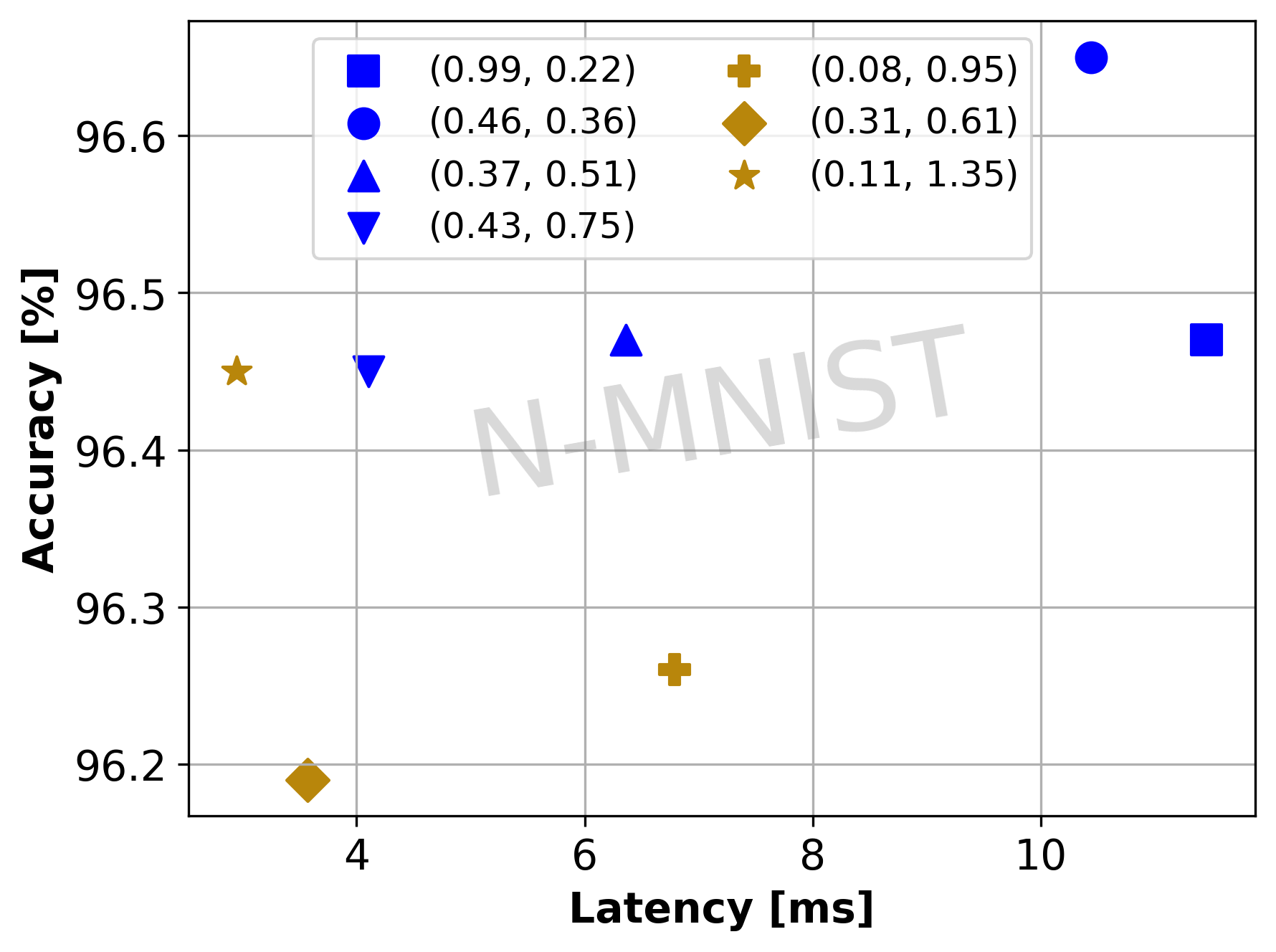} 
        \caption{}
        \label{fig:nmnist_pareto}
    \end{subfigure}
    \caption{\textbf{Pareto analysis of neuron configurations across datasets.} Each plot shows accuracy versus latency for different $(\beta, \theta)$ parameter pairs using either \textit{LIF} (blue markers) or \textit{LAP} (yellow markers). Optimal configurations appear toward the upper-left (high accuracy, low latency). \textit{LAP} consistently yields lower latency and often competitive or superior accuracy compared to \textit{LIF}.}
    \vspace{-10pt}
    \label{fig:beta_thr_sweep}
\end{figure*}

\subsection{Neuronal Model: Arithmetic Intensity vs. Sparsity}
\label{sec:results-neuron}

Figure~\ref{fig:beta_thr_sweep} visualizes the design space exploration of neuron models, plotting the Pareto frontier of accuracy vs. latency. Each point in the plot corresponds to a specific neuron model (\textit{LIF}--blue markers or \textit{LAP}--yellow markers) and its associated parameters $(\beta, \theta)$. This experiment tests a fundamental hardware hypothesis: \textit{Does the increased arithmetic intensity of complex neuron models (Lapicque) pay off in reduced event rates?}

Counterintuitively, the \textit{LAP} model, which requires more complex state updates (Eq.~\ref{eq:lap}), consistently yields lower system latency than the simpler LIF model. In DVS-Gesture (Fig.~\ref{fig:gest_pareto}), the \textit{LAP} configurations cluster in the top-left quadrant (high accuracy, low latency), whereas \textit{LIF} configurations are scattered with significantly higher latency tails ($>$ 60 ms). The optimal \textit{LAP} configuration reduces latency to $\sim$5.8 ms compared to the best \textit{LIF} result under the evaluated settings. On N-MNIST (Fig.~\ref{fig:nmnist_pareto}), \textit{LAP} provides a substantial \textbf{28\% latency reduction} at comparable ($\sim$96.5\%) accuracy.

\noindent\textbf{Analysis:} This validates that, within our training and hardware framework, the \textit{LAP} model's explicit modeling of RC constants can lead to more efficient temporal encoding and reduced event activity (i.e., fewer spikes) compared to \textit{LIF}. As illustrated in Fig.~\ref{fig:neuron_dataflow}, \textit{LIF} requires only 3 operations per update (accumulate, shift-based decay, compare), while \textit{LAP} requires 4 operations, including 2 explicit multiplications for the RC constants. Despite this $\sim$33\% higher per-update arithmetic cost, \textit{LAP} suppresses total spike events so effectively that it amortizes the additional computation. On N-MNIST, for example, the 28\% latency reduction implies that \textit{LAP} reduces total neuron updates by more than enough to offset its per-update overhead. From a hardware perspective, this validates the use of biologically plausible models in efficiency-constrained digital accelerators. The cost of complex update logic is compensated by the reduction in total accumulated events.

\subsection{Benchmarking Against Previous Work}
\label{sec:comparison-prior}

\begin{table*}[t]
\centering
\small
\caption{Benchmarking tuned configurations against previous work. Note: Our VGG9 network is significantly deeper (9 layers) than the 2--4 layer baselines, yet achieves competitive latency. EDP = Energy-Delay Product.}
\begin{tabular}{| c | c | c | c | c | c | c | c |} 
 \hline
 \multirow{2}{*}{\textbf{Dataset}} & \multirow{2}{*}{\textbf{Study}} & \multirow{2}{*}{\textbf{Net}} & \textbf{Acc.} & \textbf{Latency} & \textbf{Power} & \textbf{EDP} & $\textbf{F}_{Max}$ \\  
 &&&[\%]&[ms]&[W]&[mJ$\cdot$ms]&[MHz]\\
  \hline
  \hline
 \multirow{2}{*}{NMNIST} & ASIC \cite{di2022sne} & \hspace{0pt}net$^{1}$ & 97.8 & 3.8-12.5 & 0.011 & 0.16--1.72 & 400 \\
 & \textbf{\texttt{TW}} & \hspace{0pt}net$^{3}$ & 96.4 & 5.9 & 0.31 & 10.8 & 100 \\ 
   \hline
\multirow{3}{*}{DVS} & ASIC\cite{di2022sne} & \hspace{0pt}net$^{1}$ & 92.4 & 7.1-23.2  & 0.011 & 0.55--5.92 & 400 \\
\multirow{3}{*}{Gesture} & FPGA \cite{liu2023low} & \hspace{0pt}net$^{2}$ & 86.3 & 0.46 & 0.42 & 0.09 & 100 \\ 
 & \textbf{\texttt{TW}} & \hspace{0pt}net$^{3}$  & 95.4 & 11.2 & 1.35 & 169.3 & 100 \\ 
 & \textbf{\texttt{TW}}* & \hspace{0pt}net$^{3}$ & 95.4 & 5.8 &  2.19 & 73.7 & 100 \\ 
 \hline
\end{tabular} \\
\label{table:prev_work}
\hspace{0pt}net$^{1}$: MP4-2C3-MP2-32C3-MP2-FC512-FC11. \\
\hspace{0pt}net$^{2}$: MP4-32C3-MP2-32C3-MP2-FC512-FC256-FC11. \\
\hspace{0pt}net$^{3}$: A VGG9 variant (Section \ref{sec:network}) customized for event datasets using Optuna. \\
\textbf{\texttt{TW}}: This Work. \textbf{\texttt{TW}}*: This Work with scaled up resources. 
\vspace{-15pt}
\end{table*}

To contextualize our characterization, we benchmark our hyperparameter-tuned VGG9 models against prior ASIC and FPGA implementations. Specifically, we benchmark the optimized configurations against two notable works: an ASIC-based accelerator from \cite{di2022sne} and an FPGA implementation from \cite{liu2023low}. This comparison helps validate that improvements derived from surrogate and neuron model tuning translate into meaningful system-level gains in latency and accuracy, even when deployed on more resource-intensive networks.

Table~\ref{table:prev_work} summarizes the accuracy, inference latency, power consumption, and clock frequency ($F_\text{Max}$) for each design on the N-MNIST and DVS-Gesture datasets. We exclude DVS-CIFAR10 from this comparison due to the absence of publicly available hardware benchmarks for this dataset. Our models are built on a VGG9-based architecture (\texttt{net$^3$}), selected using Optuna and significantly larger than those used in prior studies (\texttt{net$^1$} and \texttt{net$^2$}). 

Note that Table~\ref{table:prev_work} is not intended as an apples-to-apples architectural comparison, as the platforms, network depths, and design goals differ substantially. Rather, it demonstrates a key performance insight that through sparsity-aware hyperparameter tuning alone, a deeper VGG9 network on FPGA can match or exceed the latency of untuned, shallower models on custom ASICs. This highlights that training-time hyperparameters are a first-order determinant of hardware performance, complementary to architectural optimizations.

\noindent\textbf{Latency competitiveness:} Despite using a significantly larger network (VGG9 vs. shallow CNNs in prior work) and running on a low-frequency FPGA fabric (100 MHz), our optimized configurations achieve latencies competitive with custom ASICs (400 MHz). On \textbf{N-MNIST}, our tuned configuration (surrogate=\texttt{FS}, neuron=\textit{LAP}, $\beta=0.11$, $\theta=1.35$) achieves an inference latency of \textbf{5.9 ms}, rivaling the 3.8--12.5 ms range of the ASIC in \cite{di2022sne}. On DVS-Gesture, our scaled design (\texttt{TW*}) achieves 5.8 ms, a 2$\times$ improvement over the untuned baseline (\texttt{TW}) and faster than the reported 7.1 ms of the ASIC.

\noindent\textbf{Accuracy-power trade-off:} Our approach prioritizes algorithmic robustness, achieving \textbf{95.4\% accuracy} on DVS-Gesture compared to 86.3\%. While this necessitates a larger model and consequently higher power consumption (2.19 W on FPGA vs. mW-range on ASIC), the results demonstrate that sparsity-aware tuning is a first-order optimization vector. These results indicate that carefully tuned SNN configurations can significantly improve latency on FPGA-based implementations, in some cases approaching the performance range reported for custom neuromorphic ASIC systems. 

\noindent\textbf{Energy-delay product analysis:} To provide a more holistic efficiency metric, we report Energy-Delay Product (EDP) in Table~\ref{table:prev_work}. While the absolute EDP of our FPGA implementation exceeds the mW-range ASICs due to the underlying substrate, our sparsity-aware tuning achieves a \textbf{2.3$\times$ reduction in EDP} (from 169 mJ$\cdot$ms to 74 mJ$\cdot$ms) when scaling resources on the same platform. This relative improvement is platform-independent: the sparsity-driven latency gains translate directly into proportional energy savings. On custom silicon operating at mW-level power, these same algorithmic optimizations would yield substantial absolute EDP reductions, reinforcing that hyperparameter tuning is orthogonal to---and composable with---platform-level optimizations.

These comparisons reinforce the broader message of this work: by systematically tuning training-time hyperparameters such as surrogate gradient functions and neuron parameters, we can produce SNN models that achieve competitive or superior accuracy while yielding favorable hardware performance. This demonstrates the tangible benefits of algorithm–hardware co-design in the deployment of SNNs on real-world neuromorphic platforms.

\section{Conclusion}
\label{sec:conclusion}

This work presented a systematic workload characterization quantifying the impact of training-time hyperparameters on the physical execution efficiency of Spiking Neural Networks. By bridging the abstraction gap between algorithmic definitions (surrogate gradients, neuron models) and hardware metrics (latency, activity density), we demonstrated that software-centric design choices have profound, often non-intuitive consequences for neuromorphic deployment.

Our analysis yielded three critical insights. First, surrogate gradient functions are not merely convergence tools but are deterministic predictors of hardware sparsity. Specifically, we identified that functions with exponential tails (e.g., Spike Rate Escape) can aggressively suppress latency-inducing noise in dynamic workloads like DVS-Gesture, provided their stability cliffs are carefully navigated. Second, we challenged the assumption that simpler neuron models yield faster hardware. Our results show that the \textit{Lapicque} model, despite its higher arithmetic intensity, amortizes its computational cost by encoding information more sparsely, reducing system-level latency by up to 28\% compared to standard LIF models. Third, we validated that these gains are realizable on general-purpose FPGA fabrics, where our sparsity-aware tuning allowed a VGG9-based SNN to match the latency of custom ASICs while maintaining higher accuracy.

These findings suggest that the current SNN design paradigm, which often decouples training accuracy from inference efficiency, is insufficient. We conclude that activation sparsity should be treated as a tunable hyperparameter rather than a fixed emergent property. Our future work will focus on integrating these hardware-derived sparsity cost functions directly into the training loss loop, moving from hardware-in-the-loop evaluation to hardware-aware loss formulation, ultimately enabling the automated discovery of the optimal Pareto frontier for edge neuromorphic intelligence.

\section*{Acknowledgment}
This work was partially supported by NSF Grants 1844952 and 2425567.

\bibliographystyle{IEEEtran}
\bibliography{refs}

\newpage
\appendix
\section{Artifact Appendix}

\subsection{Abstract}

This appendix describes the artifact that contains the software and hardware implementations used to produce the results reported in our paper on SNN hyperparameter characterization. The artifact includes: (1) Python training scripts for spiking neural networks with configurable surrogate gradient functions (Fast Sigmoid, Arctangent, Spike Rate Escape, Stochastic Spike Operator) and neuron models (LIF, Lapicque); (2) a cycle-accurate FPGA instrumentation platform implemented in SystemVerilog for measuring inference latency and spike counts; and (3) weight extraction utilities for deploying trained models to hardware. The artifact enables reproduction of all key results, including accuracy vs.\ slope curves (Figure~5), latency characterization (Figure~6), Pareto analysis of neuron configurations (Figure~7), and the benchmark comparisons in Table~II.

\subsection{Artifact Check-list (Meta-information)}

{\small
\begin{itemize}
  \item {\bf Algorithm:} Surrogate gradient descent for SNNs; Design space exploration with Optuna
  \item {\bf Program:} Python training scripts, SystemVerilog RTL for FPGA simulation
  \item {\bf Compilation:} Python 3.11, PyTorch 2.2.2; Xilinx Vivado \texttt{2023.2}
  \item {\bf Model:} VGG9-based convolutional SNN (4-bit quantized weights)
  \item {\bf Data set:} DVS128-Gesture, N-MNIST, DVS-CIFAR10 (via tonic library)
  \item {\bf Run-time environment:} Windows 11, CUDA 12.8
  \item {\bf Hardware:} NVIDIA GPU for training \texttt{RTX 4090}; Xilinx Kintex UltraScale+ FPGA \texttt{xcvu9p-flga2104-2L-e} for hardware validation
  \item {\bf Metrics:} Classification accuracy (\%), inference latency (ms), spike counts, Energy-Delay Product (mJ$\cdot$ms)
  \item {\bf Output:} Trained model checkpoints, extracted weights, \texttt{cycles\_and\_spikes.txt} with latency measurements
  \item {\bf Experiments:} Surrogate sweep (4 functions $\times$ slope range), neuron model sweep (LIF/LAP $\times$ $\beta$/$\theta$ grid), hardware profiling
  \item {\bf How much disk space required (approximately)?:} $\geq$ 100 GB
  \item {\bf How much time is needed to prepare workflow (approximately)?:} 1--2 hours (environment setup, dataset download)
  \item {\bf How much time is needed to complete experiments (approximately)?:} \texttt{$\sim$48 hours} for full DSE; \texttt{$\sim$20 minutes} for hardware simulation per configuration
  \item {\bf Publicly available?:} Yes
  \item {\bf Code licenses (if publicly available)?:} MIT License
  \item {\bf Data licenses (if publicly available)?:} DVS-Gesture (CC BY 4.0), N-MNIST (see IBM license), DVS-CIFAR10 (see original license)
  \item {\bf Workflow automation framework used?:} Optuna for hyperparameter optimization
  \item {\bf Archived (provide DOI)?:} \url{https://zenodo.org/records/18893738}
\end{itemize}
}

\subsection{Description}

\subsubsection{How to Access}

The artifact is publicly available at:
\begin{center}
\url{https://github.com/githubofaliyev/SNN-DSE/tree/ISPASS26}
\end{center}

An archived version with DOI is available at:
\begin{center}
\url{https://zenodo.org/records/18893738}
\end{center}

\subsubsection{Hardware Dependencies}

\begin{itemize}
  \item \textbf{Training:} NVIDIA GPU with CUDA 12.8 support (tested on \texttt{RTX 4090}). 
  \item \textbf{Hardware simulation:} Xilinx Kintex UltraScale+ FPGA (\texttt{xcvu9p-flga2104-2L-e}).
  \item \textbf{Hardware requirements:}
\begin{itemize}
    \item \textbf{Disk space:} $\geq$100~GB (datasets: 15~GB, Vivado: 70~GB, CUDA/Python: 10~GB, working files: 5~GB)
    \item \textbf{GPU memory:} $\geq$8~GB VRAM (16~GB recommended)
    \item \textbf{System RAM:} $\geq$16~GB (32~GB recommended for Vivado synthesis)
\end{itemize}
\end{itemize}

\subsubsection{Software Dependencies}

\begin{itemize}
  \item Python 3.11
  \item PyTorch 2.2.2 with CUDA 12.8
  \item snnTorch 0.7.0
  \item Brevitas 0.10.2 (for quantization-aware training)
  \item tonic (for DVS event-based datasets)
  \item Optuna (for hyperparameter optimization)
  \item Xilinx Vivado \texttt{2023.2} (for hardware simulation)
\end{itemize}

All Python dependencies are listed in \texttt{Scripts/requirements.txt}.

\subsubsection{Data Sets}

The following event-based datasets are used (automatically downloaded via tonic):
\begin{itemize}
  \item \textbf{DVS128-Gesture:} 11 gesture classes, 2 channels, 128$\times$128 resolution
  \item \textbf{N-MNIST:} 10 digit classes, 2 channels, 34$\times$34 resolution
  \item \textbf{DVS-CIFAR10:} 10 object classes, 2 channels, 128$\times$128 resolution
\end{itemize}

\subsubsection{Models}

VGG9-based convolutional SNN architecture:
\begin{verbatim}
64C3-28C3-MP2-48C3-54C3-MP2-120C3-126C3
-140C3-MP2-FC216-FC200
\end{verbatim}
Weights are quantized to 4-bit integers for hardware deployment.

\subsection{Installation}

\begin{enumerate}
  \item Clone the repository:
\begin{verbatim}
git clone https://github.com/
githubofaliyev/SNN-DSE.git
cd SNN-DSE
git checkout ISPASS26
\end{verbatim}

  \item Create and activate a Python environment:
\begin{verbatim}
conda create -n snn-dse python=3.11
conda activate snn-dse
\end{verbatim}

  \item Install Python dependencies:
\begin{verbatim}
pip install -r Scripts/requirements.txt
\end{verbatim}

  \item (For hardware simulation) Install Xilinx Vivado and ensure it is in your PATH.
\end{enumerate}

\subsection{Experiment Workflow}

The workflow consists of three phases:

\subsubsection{Phase 1: Training with Surrogate/Neuron Sweeps}

\begin{enumerate}
  \item Edit \texttt{Scripts/Configs.py} to set the desired configurations.

  E.g., 
  \begin{verbatim}
config = {
    "beta": 0.15, # Decay rate
    "threshold": 0.5, # Threshold
    "slope": 1.0, 
    # Scaling factor (alpha)
    "surrogate_type": "fast_sigmoid",  
    # "fast_sigmoid", "atan", 
    # "spike_rate_escape", "SSO"
    "neuron_type": "lif", 
    # "lif" or "lapicque"
    ...
}
  \end{verbatim}

  \item Run training:
\begin{verbatim}
cd Scripts
python Training.py
\end{verbatim}

  \item To reproduce the full DSE (Figures 5--7), sweep over:
  \begin{itemize}
    \item Surrogate functions: \texttt{fast\_sigmoid}, \texttt{atan}, \texttt{spike\_rate\_escape}, \texttt{SSO}
    \item Slope $\alpha \in [1, 48]$
    \item Neuron types: \texttt{lif}, \texttt{lapicque}
    \item $\beta \in [0.1, 1.0]$, $\theta \in [0.1, 2.0]$ with step 0.2
  \end{itemize}
\end{enumerate}

\subsubsection{Phase 2: Weight Extraction}

\begin{enumerate}
  \item Edit \texttt{Scripts/Extract.py} to specify the trained model path and dataset.
  \item Run extraction:
\begin{verbatim}
python Extract.py
\end{verbatim}
  \item Copy the generated \texttt{`include} line into \texttt{Hardware/ispass\_sim/top\_wrapper.sv}.
\end{enumerate}

\subsubsection{Phase 3: Hardware Simulation}

\begin{enumerate}
  \item Open the project in Vivado.
  \item Add \texttt{-d SIM} to \texttt{xsim.compile.xvlog.more\_options} in simulation settings.
  \item Run behavioral simulation.
  \item Simulation completes when \texttt{fc\_2\_spk\_RAM\_loaded} triggers.
  \item Results are written to \texttt{cycles\_and\_spikes.txt}.
\end{enumerate}

\subsection{Evaluation and Expected Results}

\subsubsection{Key Claims to Verify}

\begin{enumerate}
  \item \textbf{Figure 5 (Accuracy vs.\ Slope):} Fast Sigmoid maintains $>$90\% accuracy across a wide slope range on N-MNIST and DVS-Gesture; SRE exhibits cliff-like degradation.
  
  \item \textbf{Figure 6 (Latency Characterization):} On DVS-Gesture, SRE achieves 12.2\% lower latency than FS ($\sim$92 ms vs.\ $\sim$105 ms at P=1).
  
  \item \textbf{Figure 7 (Pareto Analysis):} Lapicque configurations cluster in the high-accuracy, low-latency quadrant; LAP provides up to 28\% latency reduction on N-MNIST.
  
  \item \textbf{Table II (Benchmarking):} Tuned VGG9 on FPGA achieves 5.9 ms latency on N-MNIST (comparable to ASIC range of 3.8--12.5 ms) and 95.4\% accuracy on DVS-Gesture.
\end{enumerate}

\subsubsection{Expected Output Format}

Sample: \url{https://github.com/githubofaliyev/SNN-DSE/blob/ISPASS26/Hardware/cycles_and_spikes_sample.txt} 

\subsection{Experiment Customization}

\begin{itemize}
  \item \textbf{New surrogate functions:} Add to \texttt{Net.py} using snnTorch's \texttt{surrogate} module.
  \item \textbf{Different network architectures:} Modify \texttt{Net.py} and update extraction in \texttt{Extract.py}.
  \item \textbf{New datasets:} Add dataset class to \texttt{Datasets.py} following the existing patterns.
  \item \textbf{Hardware parallelism:} Adjust the parallelization factor $P$ in the RTL configuration.
\end{itemize}

\subsection{Notes}

The full DSE involves $\sim$8,000 training epochs per surrogate function. For quick validation, we recommend testing a subset of configurations (e.g., top-2 slopes per surrogate as shown in Figure 6).

\end{document}